\documentclass[preprint2]{emulateapj}
\usepackage{color}
\usepackage{amsmath}
\usepackage{amsbsy}
\usepackage{threeparttable}
\usepackage{natbib}
\usepackage{subfigure}
\usepackage{times}
\usepackage{url}

\DeclareGraphicsExtensions{.jpg,.pdf,.png,.eps,.ps}

\newcommand{\nhat}{\hat{\mathbf{n}}}
\newcommand{\lvec}{\mathbf{l}}
\newcommand{\Lvec}{\mathbf{L}}
\newcommand{\dhat}{\hat{d}}
\newcommand{\clphi}{C_L^{\phi\phi}}

\newcommand{\dtwol}[1]{\frac{d^2 \lvec_{#1}}{(2\pi)^2}}

\newcommand{\nlzero}{\ensuremath{N_L^{(0)}}}
\newcommand{\nlone}{\ensuremath{N_L^{(1)}}}

\newcommand{\sate}{\ensuremath{\sigma_8}}

\newcommand{\ac}{\ensuremath{A_c}}

\providecommand{\e}[1]{\ensuremath{\times 10^{#1}}}
\newcommand{\Msol}{\ensuremath{\rm M_{\odot}}}
\newcommand{\msol}{\ensuremath{\rm M_{\odot}}}

\newcommand{\lmax}{\ensuremath{ l_{\rm max}}}
\newcommand{\smax}{\ensuremath{ S_{\rm max}}}

\def\ell{l}

\newcommand{\trisp}{\ensuremath{\mathcal{T}}}
\newcommand{\mvir}{\ensuremath{M_{\rm{vir}}}}

\newcommand{\Planck}{{{\it Planck}}}

\def\Stonybrook{1}
\def\Argonne{2}

\def\McGill{3}

\def\BCCP{4}
\def\Yale{5}

\begin{document}

 \title{CMB lensing power spectrum biases from galaxies and clusters \\using high-angular resolution temperature maps}

\vskip 10mm

\author{A. van Engelen\altaffilmark{\Stonybrook}, 
  S.~Bhattacharya\altaffilmark{\Argonne}, N.~Sehgal\altaffilmark{\Stonybrook}, G.~P.~Holder\altaffilmark{\McGill},
  O.~Zahn\altaffilmark{\BCCP}, D.~Nagai\altaffilmark{\Yale}}

\altaffiltext{\Stonybrook}{Department of Physics and Astronomy,
Stony Brook University, Stony Brook, NY, 11794}

\altaffiltext{\McGill}{Department of Physics,
McGill University, 3600 Rue University, 
Montreal, Quebec H3A 2T8, Canada}

\altaffiltext{\Argonne}{Argonne National Laboratory, 9700 S. Cass Avenue, Argonne, IL, USA 60439}

\altaffiltext{\BCCP}{Berkeley Center for Cosmological Physics,
Department of Physics, University of California, and Lawrence Berkeley
National Labs, Berkeley, CA, USA 94720}
\altaffiltext{\Yale}{Department of Physics, Department of Astronomy and Yale Center for Astronomy \& Astrophysics, Yale University, New Haven, CT 06520}

\begin{abstract}
  The lensing power spectrum from cosmic microwave background (CMB) temperature maps will be measured with unprecedented precision with upcoming  experiments, including upgrades to ACT and SPT.   Achieving significant improvements in cosmological parameter constraints, such as percent level errors on $\sigma_8$ and an uncertainty on the total neutrino mass of $\sim50$ meV, requires percent level measurements of the CMB lensing power. This necessitates tight control of systematic biases.   We study several types of biases to the temperature-based lensing reconstruction signal from foreground sources such as radio and infrared galaxies and the thermal Sunyaev-Zel'dovich effect from galaxy clusters.  These foregrounds bias the CMB lensing signal due to their non-Gaussian nature.  Using simulations as well as some analytical models  we find that these sources can substantially impact the measured signal if left untreated.  
However, these biases can be brought to the percent level if one masks galaxies with fluxes at 150 GHz above 1~mJy and galaxy clusters with masses above $M_{\rm{vir}} = 10^{14} M_{\odot}$.  To achieve such percent level bias, we find that only modes up to a maximum multipole of  $l_{\rm{max}} \sim 2500$ should be included in the lensing reconstruction.  We also discuss ways to minimize additional bias induced by such aggressive foreground masking by, for example, exploring a two-step masking and in-painting algorithm. 
\end{abstract}
\vskip 5mm

\section{Introduction}

Gravitational lensing of the cosmic microwave background (CMB) by
intervening large-scale structure has long been recognized as a powerful probe of cosmology (see \citealt{lewis06} for a theoretical review).  Only recently, however, have millimeter-wave experiments  reached the sensitivity to detect this promising signal.

There are two main methods for detecting CMB lensing.  One is to measure the smoothing of
the acoustic peaks on small angular scales induced by lensing in
the CMB power specrum \citep{seljak96,challinor05}.  This has been detected using data from the 
Arcminute Cosmology Bolometer Array Receiver (ACBAR;
\citealt{reichardt09a}), the Atacama Cosmology Telescope (ACT;
\citealt{dunkley11, das13}), the South Pole Telescope (SPT;
\citealt{keisler11, story12}), and the {\it Planck} satellite (\citealt{planck13_p16}).

The second method is to measure the distinctive
mode-coupling that lensing generates in the CMB.  
Optimal filters exploiting this mode coupling can be applied to millimeter-wave maps 
to generate reconstructed maps of the matter distribution responsible for the lensing
\citep{bernardeau97,seljak99, zaldarriaga99, hu01b}.  The power spectra of these reconstructed 
maps measure the matter power spectrum integrated along the line of sight.  This method has the advantages of ultimately allowing one to measure the lensing power spectrum directly as a function of scale, and additionally of providing a higher significance detection.
Reconstructions done by ACT \citep{das11b, das13} and SPT
\citep{vanengelen12}, yielded lensing detection significances
of $4$ and $6.3\sigma$ respectively.  More recently {\it Planck } \citep{planck13_p17} has achieved a detection significance of $25\sigma$. 

To date, all reconstructed lensing maps obtained from data have employed the optimal
quadratic estimation technique of \citet{hu01b} and \citet{hu02a}.
This technique is derived theoretically under idealized
experimental conditions, in which the CMB is taken to be a nearly
perfect Gaussian field with only a small mode-coupling induced by
gravitational lensing, and with no significant mode-coupling from instrumental or
foreground effects.  Since then, the realities of data have demanded 
the investigation of a number of additional sources of mode coupling, including 
 finite sky coverage
\citep{perotto10, vanengelen12, namikawa12, benoitlevy13}, nonstationary noise statistics 
\citep{hanson09}, primordial non-Gaussianity \citep{lesgourgues05,merkel13}, and higher-order mode couplings induced by lensing itself \citep{kesden03, hanson11}.

In addition, astrophysical foregrounds in CMB maps intrinsically have significant mode coupling due to their non-Gaussian nature.  
\citet{amblard04} estimated levels of
contamination based on simulations of large-scale structure, finding
biases on the reconstructed lensing power spectrum ranging 
between tens of percents to factors of two due to the presence
of thermal and kinetic Sunyaev-Zel'dovich effects.  In their
detections of the CMB lensing power spectrum, 
\citet{das11b}, \citet{vanengelen12}, and \citet{das13} argued that the
effects of astrophysical foregrounds could be neglected at  sensitivity
levels presented in those works.
This was based partly on an improved
understanding of the properties of the millimeter-wave sky as measured by ACT and SPT
since the publication of \citet{amblard04}.
These improvements include more accurate measurements of the power spectrum amplitudes of point sources and the thermal and kinetic SZ effects \citep{hall10, dunkley11,shirokoff11,reichardt11}.  
The biasing signals from
astrophysical sources were also minimized in the SPT analysis by filtering out the smallest CMB scales, which minimized the non-Gaussian foreground structure in the maps.  
In the recent strong detection of lensing by the Planck collaboration, the lensing signal was  measured 
using even larger CMB scales, due to the resolution of the instrument, such that biases were negligible apart from a small term associated with rare bright sources.  This was a $\sim 2$\% positive bias that was subtracted off from power spectrum estimates \citep{planck13_p17}.

Measures of the lensing power spectrum are expected to continue to
improve in signal-to-noise ratio.  The lensing analysis of the full SPT temperature
survey, which consists of about four times the survey area considered in
\citet{vanengelen12}, is ongoing.  In addition, ACT and SPT have both
been upgraded with polarization-sensitive receivers \citep{niemack10,mcmahon09}, and the new instruments (ACTpol and SPTpol) are currently taking data.  
ACTpol, in particular, will survey much wider areas than achieved to date with ground-based CMB instruments, and is expected to yield lensing detections in temperature maps that will roughly double the signal-to-noise ratio.  This will allow statistical uncertainty in the lensing power spectrum to reach the $\sim$$2\%$ level, improving current errors by a factor of two \citep{niemack10}.  Achieving this will require biases 
to the lensing power spectrum from astrophysical foregrounds to be constrained at the percent level.  

In this paper, we quantify the astrophysical biases with
an eye toward these upcoming datasets.  We consider the contributions
from millimeter-wave emission by infrared galaxies and radio galaxies, as well as the contribution from galaxy clusters via the thermal
Sunyaev-Zel'dovich effect.  We focus this work on biases to lensing reconstruction from temperature maps, deferring the study of foreground bias in polarization-based reconstructions to future work.  In all that follows, we work in the flat-sky approximation, where harmonic transforms on the sphere are approximated using Fourier methods.


\section{CMB Lensing Reconstruction}
In the absence of lensing and other sources of CMB mode-coupling, the
CMB is a globally Gaussian field.  On a given scale $l$, it is fully described by its
power spectrum  according to
\begin{equation}  
  \label{eq:diagcoupling}
  \langle
  T^U(\lvec_1) T^U(\lvec_2) \rangle = (2\pi)^2 \delta(\lvec_1 + \lvec_2)
  C_{\ell_1}^U. 
\end{equation}
Here, $T^U(\lvec)$ is the Fourier transform of the unlensed CMB map, and $C_\ell^U$ is the unlensed power spectrum.
Lensing shifts the unlensed CMB temperature at position $\nhat$, to a new position, $\nhat + {\boldsymbol{\alpha}}$, where the deflection angle ${\boldsymbol{\alpha}}$
is given by the gradient of a scalar field $\phi(\nhat)$.  Thus
\begin{align} 
  T(\nhat) =& T^U(\nhat + \nabla \phi(\nhat)).
  \label{eq:lensingformula}
\end{align}
This scalar field is the weighted line-of-sight
projection of the three-dimensional gravitational potential of matter
between the observer and the CMB, and thus probes the evolution of cosmic structure through time.

Since the deflection angle represents a small perturbation about $\nhat$, we can Taylor expand, which to first order gives  
\begin{align}
T(\lvec) =T^U(\lvec) + (\nabla T^U \star \nabla \phi)(\lvec),
\end{align}
where pixel-by-pixel multiplication becomes convolution in Fourier space.
Since $\nabla T(\lvec) = -i\lvec T(\lvec)$, then for $\Lvec = \lvec_1 + \lvec_2$ and $\Lvec \ne 0$, keeping only terms linear in $\phi$ yields
\begin{align} \label{eq:lensmodecoupling}
  \langle T(\lvec_1) T(\lvec_2) \rangle_{\rm CMB} = & (\lvec_1 + \lvec_2) \cdot (\lvec_1
  C_{\ell_1}^U + \lvec_2 C_{\ell_2}^U) \phi(\lvec_1 + \lvec_2) \nonumber \\
  \equiv & f(\lvec_1, \lvec_2) \phi(\Lvec).
\end{align}
We choose to focus on the $\Lvec \neq 0$ terms because the $\Lvec = 0$ terms led to the power spectrum.  
The subscript ``CMB'' on the left-hand side indicates that we consider an ensemble average of CMB realizations, while holding the $\phi$ realization fixed.   Eq.  \ref{eq:lensmodecoupling} describes the mode coupling lensing generates.  If there was no lensing, i.e. $\phi = 0$, then this would be zero.

The optimal quadratic estimator for the lensing field $\phi$ is formulated to take advantage of this mode coupling \citep{hu00, hu01b}.  This is done by filtering the maps with a filtering function, $F(\lvec_1, \lvec_2)$, that downweights the noisy modes, and selects for the mode-coupling of
Eq.~\ref{eq:lensmodecoupling}:
\begin{equation}
  F(\lvec_1, \lvec_2) = {f(\lvec_1, \lvec_2) \over 2C_{l_1}^tC_{l_2}^t}.
\label{eq:filter}
\end{equation}
Here, $C_l^t$ is the assumed total power in the map, including CMB, noise, and foreground power.

Thus, we can estimate the lensing deflection field, $\dhat$, which is related to the
lensing potential via $d(\Lvec) = L\phi(\Lvec)$, using
\begin{equation}
  \dhat(\Lvec) \equiv L{\hat\phi}(\Lvec) =   \frac{A_\Lvec}{L} \int \dtwol{1} F(\lvec_1, \Lvec -
  \lvec_1) T(\lvec_1) T(\Lvec - \lvec_1). \label{eq:reconst_defl}
\end{equation}
In this reconstruction of the deflection field, the normalization $A_\Lvec$ is chosen to ensure that the
resulting lensing map is unbiased, with $\langle \dhat(\Lvec) \rangle_{\rm
  CMB} = L\phi(\Lvec)$.  

An estimation of the lensing power spectrum can be obtained from  $\dhat(\Lvec)$ by computing
\begin{align}
  \langle  \hat{d}(\Lvec_1) \hat{d}(\Lvec_2) \rangle &= {A_{\Lvec_1} \over L_1}{A_{\Lvec_2} \over L_2}  \int \dtwol{1} \int \dtwol{2} \nonumber \\
  & \times F(\lvec_1, \Lvec_1 - \lvec_1) F(\lvec_2, \Lvec_2 - \lvec_2) \nonumber \\ 
  & \times \langle T(\lvec_1) T(\Lvec_1 - \lvec_1) T(\lvec_2) T(\Lvec_2 - \lvec_2) \rangle. \label{eq:deflpower}
\end{align}
The estimated lensing power spectrum is thus sensitive to the four-point product of CMB modes in the Fourier domain.\footnote{The angled brackets   $\langle . \rangle$ without the ``CMB'' subscript indicate that this ensemble average is taken over realizations of both the $T^U(\lvec)$ and $\phi(\lvec)$ fields;  see \citet{kesden03} for a more detailed discussion.  We are neglecting the correlation between the  two fields, induced by the integrated Sachs-Wolfe effect.}
This four-point product  can be decomposed into several terms:
\begin{align}
  \langle T(\lvec_1) T(\lvec_2) T(\lvec_3) T(\lvec_4) \rangle = & ( \langle T(\lvec_1) T(\lvec_2) \rangle\langle T(\lvec_3) T(\lvec_4) \rangle \nonumber \\ 
  & +2~{\rm{perm.}})\nonumber \\ 
  & + \langle T(\lvec_1) T(\lvec_2) T(\lvec_3) T(\lvec_4) \rangle_{\rm conn.} \label{eq:fourptdef}
\end{align}
The first two lines describe unconnected terms, which exist even for a
globally Gaussian field with no lensing.  We have used the Wick theorem to write this
unconnected piece as the sum of three sets of two-point products.  The
(+2 perm.) refers to the terms obtained by replacing the pairings
$(1,2)(3,4)$ with the pairings $(1,3)(2,4)$ and $(1,4)(2,3)$.  These
unconnected terms are given in terms of  the  power spectrum via the analogue of
Eq.~\ref{eq:diagcoupling} using the total power $C_l^t$.  

The first unconnected term is proportional to a delta function at $\Lvec = 0$, giving no contribution at $\Lvec \neq 0$.  However, the two terms represented by the (+2 perm.) lead to an effective noise bias of
\begin{equation}
  { \nlzero} = \frac{A_L^2}{L^2} \int \dtwol{1} f(\lvec_1, \Lvec - \lvec_1) F(\lvec_1, \Lvec - \lvec_1)\left[ {({ C_{l_1}^t})^\prime({ C_{|\Lvec - \lvec_1|}^t})^\prime \over  C_{l_1}^t C_{|\Lvec - \lvec_1|}^t}\right ].  \label{eq:wrongnormal}
\end{equation}
Here, $({ C_{l}^t})^\prime$ represents the total power spectrum in the field,
as opposed to the  power assumed in the denominator of the optimal lensing filter, ${ C_{l}^t}$ in Eq.\ref{eq:filter}.

The \nlzero bias thus originates from the nonzero CMB and noise power in the map.  
The superscript $(0)$ refers to the bias being zeroth order in lensing,
since it is present if $\phi = 0$.  If the map contains
the same power as that assumed in the filter, 
so
that $({ C_{l_2}^t})^\prime = { C_{l_2}^t}$, then $N_\Lvec^{(0)} =
A_\Lvec$ \citep[e.g.][]{hu02a}.  In general, there will be some scatter in \nlzero from
realization to realization due to the sample variance in the observed
CMB and noise fluctuations.  This bias can be characterized by directly evaluating
Eq.~\ref{eq:wrongnormal}, taking ${C_{l}^t}^\prime$ to be the estimated power spectrum from the
given map realization \citep{dvorkinsmith09, namikawa12}.\footnote{Alternatively, the noise bias \nlzero can be avoided altogether by splitting the Fourier domain into disjoint regions such that $\lvec_1$ and $\lvec_2$ do not share modes in common with $\lvec_3$ and $\lvec_4$ \citep{sherwin10, vanengelen12}.  We do not consider this further because the characterization of the noise bias directly from the data has been shown to be tenable \citep{namikawa12, planck13_p17}.}

The connected terms in Eq. \ref{eq:fourptdef} give the trispectrum, 
\begin{align}
  \langle T(\lvec_1) & T(\lvec_2) T(\lvec_3) T(\lvec_4) \rangle_{\rm conn.} \equiv  \nonumber \\ & (2\pi)^2  \delta(\lvec_1 + \lvec_2 + \lvec_3 + \lvec_4) \trisp(\lvec_1, \lvec_2, \lvec_3, \lvec_4).
\end{align}
At first order in the lensing potential power spectrum, $\clphi$, the  trispectrum  from the lensing of the CMB is \citep{kesden03}:
\begin{align}
  \trisp(\lvec_1, \lvec_2, & \lvec_3, \lvec_4) =  
   \phantom{ + } C_{|\lvec_1 + \lvec_2|}^{\phi\phi} f(\lvec_1, \lvec_2) f(\lvec_3, \lvec_4)  \mbox{ +2 perm.}
\end{align}
The first of these terms leads to the direct measure of the CMB lensing power spectrum.  The (+2 perm.) terms lead to the \nlone~bias, which are of the same order in the lensing potential power spectrum.  This bias, and those at higher orders in $\clphi$ have been studied elsewhere \citep{kesden03,hanson11}.

Other sources of nonzero trispectrum will also lead to signals which can bias the resulting lensing power spectrum estimates. 
In particular, a field with a given trispectrum ${\trisp}^\prime(\lvec_1, \lvec_2, \lvec_3, \lvec_4)$ will lead to a response in the lensing power spectrum given by 
\begin{align}
 ({\clphi})^\prime = &\frac{A_\Lvec^2}{L^2} \int \dtwol{1}   \dtwol{2} F(\lvec_1, \Lvec - \lvec_1)  F(\lvec_2, \Lvec - \lvec_2)\nonumber \\ & \times {\trisp}^\prime(\lvec_1, \Lvec - \lvec_1, \lvec_2, \Lvec - \lvec_2). 
\label{eq:trisp_biascalc}
\end{align}  
In this work, we quantify these biases from known astrophysical sources using simulations and some analytic models.  For the majority of the paper we assume map noise levels of $18\,\mu$K-arcmin and foreground power of $9.1\,\mu$K-arcmin, which we assume to be independent of $l$.  These values are roughly consistent with expected SPT and ACTpol survey levels.   We also impose a maximum temperature multipole used in the lens reconstruction of $\lmax = 3000$.  In Section~\ref{sec:lmax}, we relax these assumptions and study the dependence of foreground biases on the map noise level and filtering choice.

\section{Simulations of the Microwave Sky}
\label{sec:simulations}
To quantify biases on the lensing power spectrum due to foregrounds we use two independent sets of simulations.  The first is described in Bhattacharya et al. 
(in preparation; B13 hereafter) and summarized below.  The second is from \citet{sehgal10} (S10 hereafter), with some modifications which we also describe below.

The B13 simulations use the Coyote N-body simulation performed using the publicly available Gadget-2 code \citep{heitmann10} to describe the dark matter distribution. The simulation has a box size of $1300$~Mpc and $1024^3$ particles. A standard $\Lambda$CDM cosmology is adopted with $\Omega_m h^2=0.1296$, $\Omega_b h^2=0.0224$, $n_s=0.97$, $\sigma_8=0.8$, $h=0.72$, and $\Omega_k=0$, consistent with the latest best-fit cosmological model from WMAP-7 \citep{Komatsu11}.   Ten simulation outputs between $z=0$ and $4$ were generated, equally spaced in the scale factor, to create a lightcone filling one octant of sky. Ray tracing through the lightcone was done to create the lensing convergence field.  The SZ effect was added to halos identified in the N-body simulation by using a semi-analytic model for gas physics \citep{shaw10}.  Infrared galaxies were also added to the halos using a semi-analytic approach. 

The flux distribution of the infrared galaxies, which constitute the cosmic infrared background (CIB), in the B13 simulations was allowed to have mass and redshift dependence in addition to frequency dependence.  To model this galaxy distribution, the dark matter halos were first modeled as NFW profiles \citep{navarro97}.  Flux from infrared galaxies was then added to these profiles following
\begin{equation}
\frac{dI}{dMdz}(M, z)= \rho_{\rm DM}(M,z) A\left(\frac{M}{M_{\rm piv}}\right)^\alpha (1+z)^\beta , 
\end{equation}
where $M$ and $z$ are the halo mass and redshift.  The halos were populated with flux using $A$, $\alpha$, and $\beta$ values with dark matter halo mass M $\ge  2.5\times 10^{12}h^{-1}M_\odot$. The smallest halo in this simulation is thus resolved by 50 particles. We consider the minimum number of parameters needed to fit the CIB clustered power spectrum \citep{reichardt12,sievers13} and the CIB bispectrum \citep{crawford13}. The advantage with this approach is that there are a few parameters with which to explore the  parameter space. The disadvantage is that one cannot predict how different galaxy populations contribute to the infrared background power.  There is also significant choice in which halos receive infrared flux.  Two such choices are made, yielding models constructed to give the same amount of clustered power seen in experiments \citep{reichardt11,sievers13}, and to match the infrared intensity distribution of the \citet{bethermin11} model.  For the CIB analysis described below, we additionally scale the CIB maps from the two models by factors of 1.21 and 1.12, to exactly match the amplitude of the clustered CIB  at 150 GHz found by \citet{reichardt11} at $l=3000$.

\begin{figure}
  \includegraphics[width=3.5in]{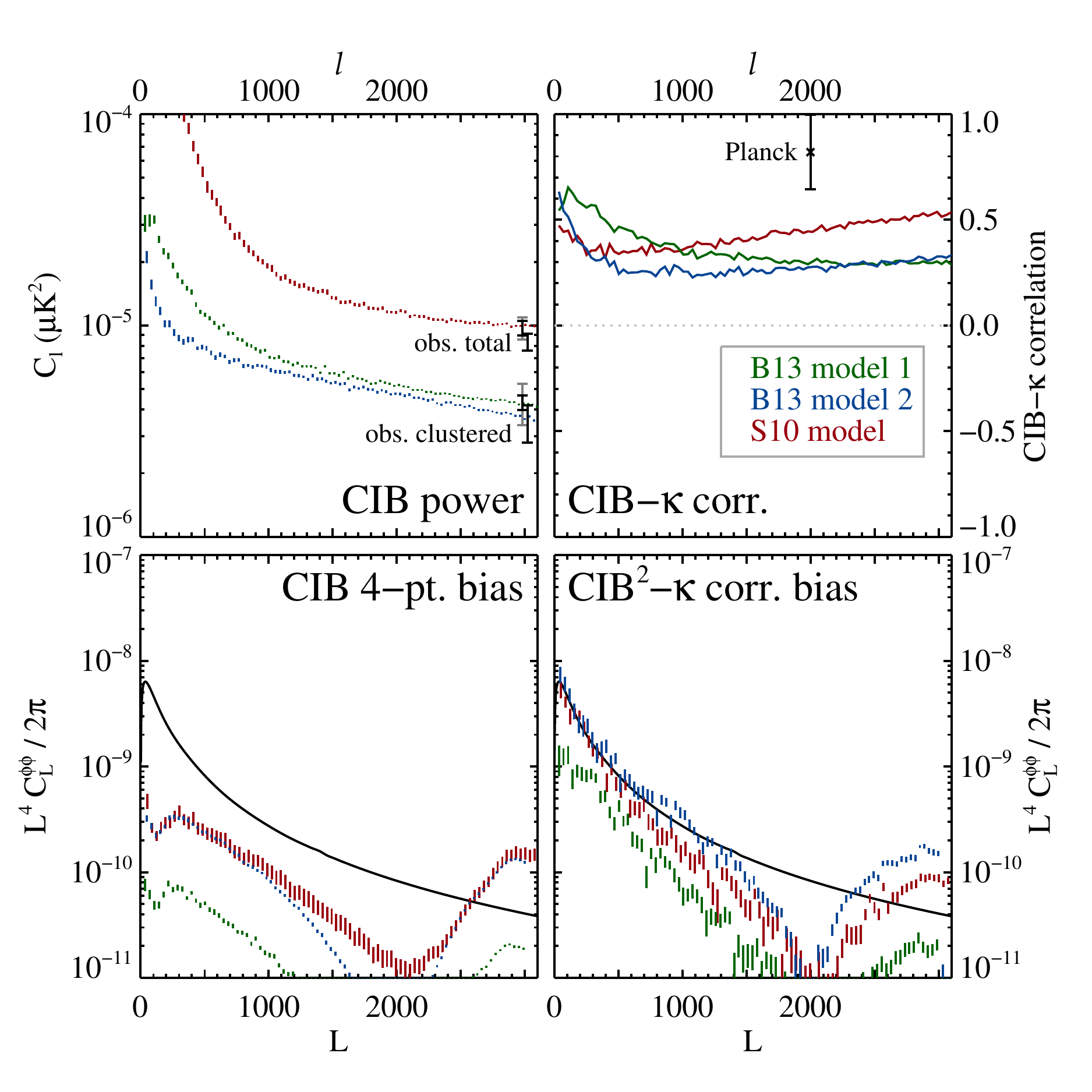}
  \caption{ Biases on lensing power spectra from the CIB. In all panels, source cuts have been applied to $5\,$mJy. Top left: power spectra of simulations used, including model 1 (green) and model 2 (blue) of the B13 simualations, together with the S10 simulations.  We also show the SPT (left black error bar, \citealt{reichardt11}) and ACT (right black error bar, \citealt{das13}) measurements. The grey error bars for SPT indicate the spread of amplitudes at $l=3000$ for the five CIB models considered by \citet{reichardt11}.     Bottom left: resulting biases from the trispectrum of the simulated sources.  Top right: CIB cross correlation with the projected mass fluctuations, $\kappa$, that lens the CMB.  We also show in black the cross-correlation found by \citet{planck13_p17}.  Bottom right: absolute value of the bias originating from correlations between the CIB and $\kappa$ fields.  The black curves show the lensing power spectrum multiplied by 0.05 for a fiducial $\Lambda {\rm CDM}$ cosmology.}
  \label{fig:cib_forpaper}
\end{figure}

The second set of simulations we use is the S10 simulations, which are described in \citet{sehgal10} and are publicly available.  We make two modifications to these simulations that differ from the original S10 version.  The first is that the SZ gas model that is implemented is instead the  gas model described in \citet{bode12}.  The second is that the fluxes of all the infrared galaxies in this simulation have been scaled down by $25\%$.   This reduction in flux makes the infrared galaxy model of S10 match the amplitude of the total infrared background power spectrum measured by ACT and SPT at 150 GHz at $l=3000$ \citep{reichardt12,sievers13}.  This infrared model also matches the source counts measured by SCUBA at 350 GHz \citep{coppin05}, the total infrared background intensity measured by FIRAS \citep{fixsen98}, and the bispectrum measured by both SPT and {\it Planck} \citep{crawford13,planck13_p30}.\footnote{Note that the S10 infrared galaxy model also predicts an abundance of infrared galaxies in massive clusters that is only 30 times larger than the field, contrary to the claims in \citet{lueker10} and \citet{hall10}.  This abundance of infrared galaxies in massive clusters is completely consistent with the measurements of \citet{bai07}.}  The S10 simulations including these two new modifications are publicly available at \url{http://www.slac.stanford.edu/$\sim$sehgal/simsv2.0/}. 

\begin{figure}[tb]
  \includegraphics[width=3.6in]{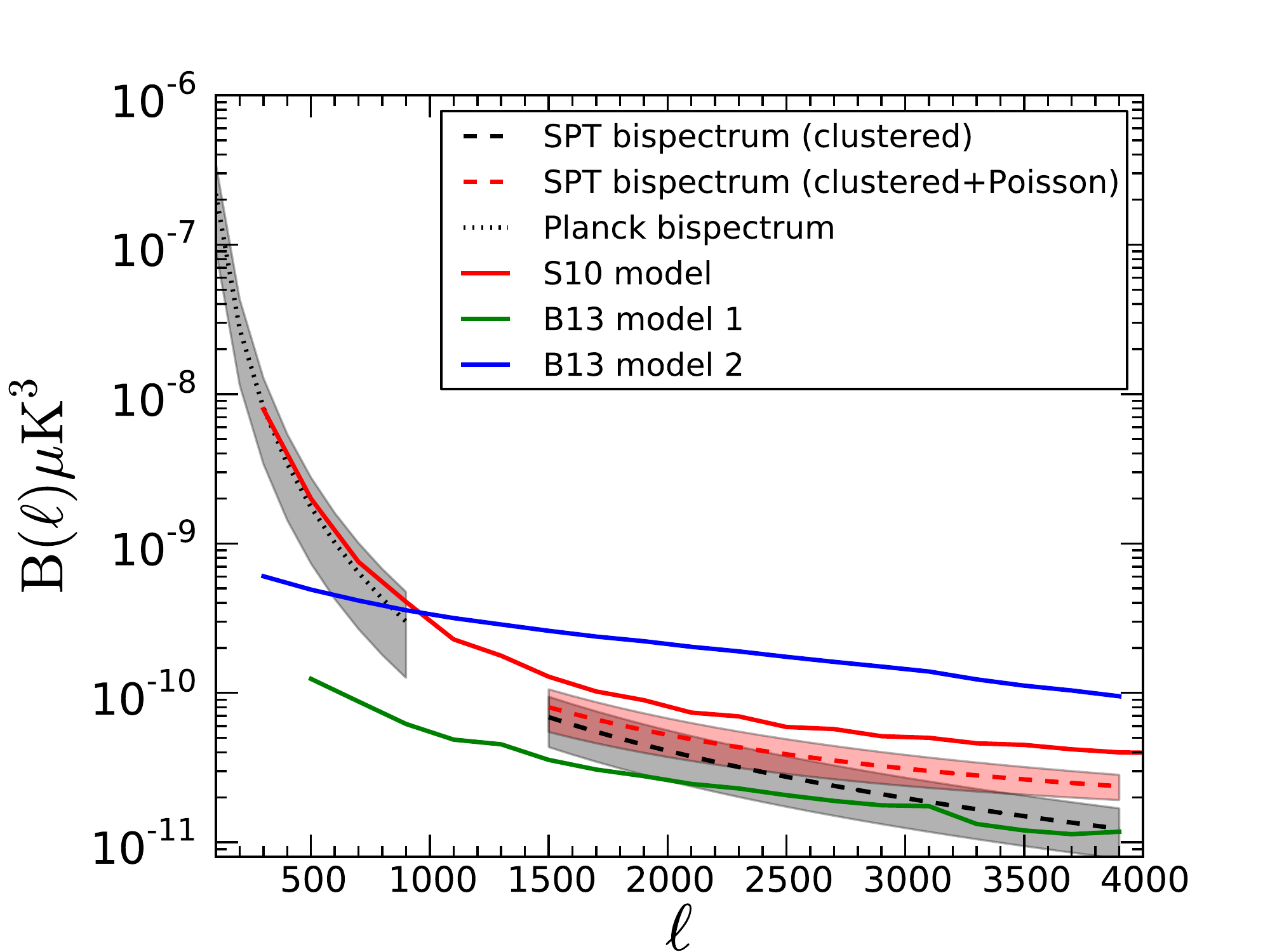}
  \caption{Bispectra of the CIB simulations, including the S10 model (red) and the two B13 models (green and blue).  The SPT  \citep{crawford13} and Planck \citep{planck13_p30} data, both scaled from 220 GHz down to 150 GHz using the procedure described in the text, are shown as the dashed and dotted curves, respectively.  The error bands include uncertainties in the frequency scaling in addition to the reported errors. \vspace{3mm}}
  \label{fig:cibbispec}
\end{figure}

The power spectra for the two CIB simulations we consider, including the flux rescalings, are shown in the upper-left panel of Fig.~\ref{fig:cib_forpaper}.  Here, sources above 5\,mJy have been removed to approximately match the 6.4\,mJy cut of the SPT analysis.\footnote{ACT removed sources above 15 mJy, which results in a negligible difference in power compared to the 6.4 mJy cut.}  We also estimate the bispectrum for our CIB models, using the tools described in detail in \citet{crawford13}, including masks to 5\,mJy.  These are shown in  Fig.~\ref{fig:cibbispec}, together with the measurements from SPT on scales $l>1500$ \citep{crawford13}, and from {\it Planck} on scales $l<800$ \citep{planck13_p30}.  For both experiments, we scale the more significant measurement  at 220\,GHz to 150\,GHz with a flux factor given by the ratio of the Poisson CIB bispectrum measurement at 220 to that at 150 GHz, which is  6.03 $\pm$ 2.23 \citep{crawford13}.  We divide the 220\, GHz clustered CIB measurements by this factor to obtain the measurement at 150 GHz. The final error bar includes uncertainty in the Poisson power at 150 and 220 \, GHz, and the uncertainty in clustered power at 220 GHz. The S10 model and the first B13 model are consistent with the SPT-obtained bispectrum. The S10 model is also consistent with the bispectrum recently measured by Planck at low multipoles ($l \leq 800$).  The second B13 model has  a larger bispectrum than that seen by SPT by a factor of $\sim 10$, indicating that lensing biases obtained from this model may possibly be overestimated.

\section{Bias on Lensing Reconstruction from Galaxies}

Emission from active galactic nuclei and dusty star-forming galaxies represent the
dominant sources of fluctuation in millimeter-wave maps on small
angular scales, and can significantly bias lensing estimates.  Below we study two different types of bias from galaxies, 
one due to the intrinsic galaxy four-point function and one due to the correlation between galaxies and the lensing field.

\subsection{Bias from the Galaxy Four-point Function}

\subsubsection{Poisson-Distributed Galaxies}
A field of uncorrelated, Poisson-distributed point sources will
possess a trispectrum.   Galaxies, in fact, do not follow a Poisson distribution since 
they are inherently correlated, residing in clustered dark matter halos.  However, we
present the scenario as an instructive toy model.  In the limit of a few bright sources, the distribution of galaxies can 
approach Poisson.

If uncorrelated sources per area element on the sky are
sufficiently numerous, the central limit theorem will apply and the
field will approach a Gaussian field with white noise.  In lensing
power spectrum reconstruction, it will then form a portion of the
$\nlzero$ bias and will be removed.  If this limit does not apply, the
uncorrelated sources will generate a trispectrum that will appear as a source of
bias in the inferred lensing power spectrum.  With high-resolution
maps from experiments such as ACT and SPT, bright, rare
sources have been cleaned to relatively faint flux thresholds to
approximately 5~mJy  \citep{das13, story12}; however, a possible
concern is the impact of sources just below the cut in flux.

A galaxy at position $\nhat$ with flux density $S$ will contribute a
temperature fluctuation of $T(\nhat) = G_\nu S \delta_D(\nhat)$ to the CMB
map, where $G_\nu$ is a factor to convert between flux density and
temperature units. This factor is given by $G_\nu = dB_\nu(T) / dT$, evaluated at the CMB temperature $T=2.73$\,K, where $B_\nu$ is the Planck function at frequency $\nu$.   In the Fourier domain, the contribution from such a galaxy corresponds to a
temperature field of
\begin{equation}
  T(\lvec) = G_\nu S e^{-i \lvec \cdot \nhat}.    \label{eq:ptsrcmodel}
\end{equation}
The power spectrum from this is constant in $\lvec$.  

For two sources, with  flux densities $S_1$ and $S_2$, at uncorrelated locations, the power spectrum is given by 
\begin{equation}
  \langle T^\star(\lvec) T(\lvec) \rangle = (G_\nu S_1)^2  + (G_\nu S_2)^2; 
\end{equation}
the cross terms cancel due to the uncorrelated nature of the sources.  This result generalizes to $N$ sources, leading to  a power spectrum for Poisson sources, $C_l^p$, of 
\begin{equation}
  C_l^p = G_\nu^2 \int_{S_{min}}^{S_{max}} dS \, S^2 {dN \over dS}. \label{eq:theoryptsrcpower}
\end{equation}
The factor ${dN \over dS}$ is the number density of sources per steradian, per unit flux interval.
Following similar arguments, the trispectrum is also constant in $\lvec$ and is given by
\begin{equation}
  \trisp(\lvec_1, \lvec_2, \lvec_3, \lvec_4) = G_\nu^4 \int_{S_{min}}^{S_{max}} dS S^4 {dN \over dS}.
\end{equation}

The  power spectrum and trispectrum of sources both lead  to signatures in reconstructed CMB lensing power spectra.  
The source power spectrum yields a portion of the  $\nlzero$ bias following Eq.~\ref{eq:wrongnormal}, and is hence removed in an analysis, while the source trispectrum  leads to an additive bias following Eq.~\ref{eq:trisp_biascalc}.

We evaluate  these expressions as a function of lensing multipole $L$ for two distinct galaxy populations. 
The number counts for extragalactic sources which appear at mm-wave
frequencies have been well-studied at the bright end, where individual
sources can be identified \citep[e.g.,][]{vieira10, marsden13, mocanu13}.  Models have been made that extrapolate and interpolate these measured counts to lower flux levels \citep{negrello07, dezotti05}.   First, we consider dusty, star-forming galaxies, for which we use the model of  \citet{negrello07}.    Since the lowest frequency treated explicitly by this model is 350~GHz (or $850\,~\mu$m),  one needs to scale the sources from 350 GHz to 150 GHz; we scale to match the predicted Poisson power spectrum of this model to the measurement obtained by the  SPT group, $l(l+1)C_l/2\pi =7.5 \pm 0.5~\,\mu$K$^2$, assuming infrared sources have a  Poisson component \citep{reichardt12}.  The SPT measurement is for a flux cut of 6.4\,mJy, and is in agreement with that of the ACT group \citep{sievers13}.  

\begin{figure}[tb]
  \includegraphics[width=3.5in]{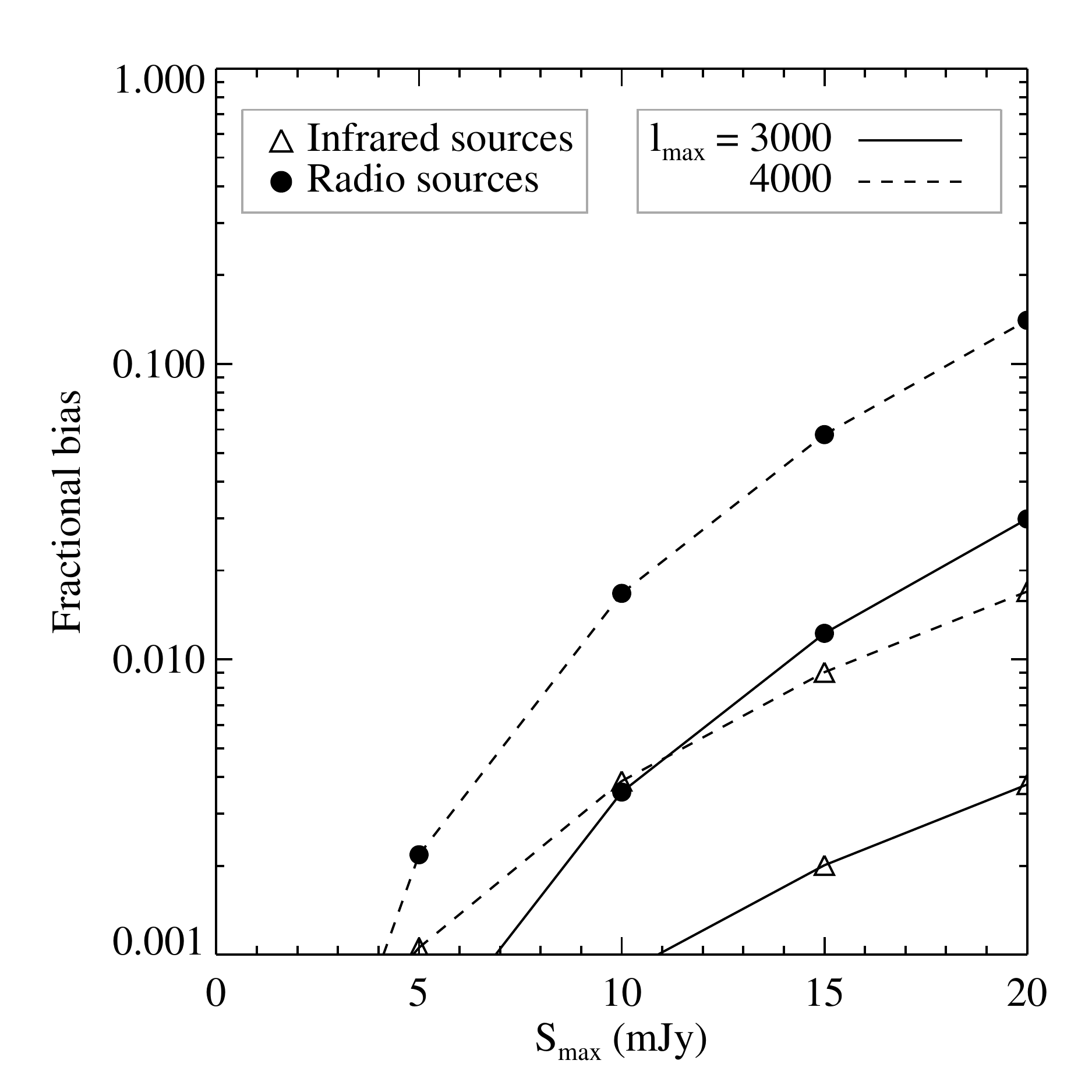}
  \caption{Fractional bias on the lensing power spectrum for Poisson-distributed point sources.  Open triangles show dusty galaxies from the \citet{negrello07} model, and solid circles show radio galaxies from the \citet{dezotti05} model.  Also shown are two difference choices of maximum multipole, $\lmax = 3000$ and $\lmax = 4000$.}
  \label{fig:poisson_vs_smax}
\end{figure}

The trispectrum term from this calculation is shown as a function of flux cut and is evaluated at the lensing multipole $L=500$ in Fig. \ref{fig:poisson_vs_smax}.  We choose $L=500$ as a representative scale because it is near the middle of the distribution of lensing information per multipole for an experiment like SPT or ACT.  The biases fall with  more aggressive flux cut, and are below percent level for  $S_{\rm max} \simeq 14\,$mJy for $\lmax = 3000$.  We have confirmed this result with lensing reconstructions of simulations of Poisson-distributed sources with the same $dN/dS$.  

The regime for ACT/SPT stands in contrast to that of the  {\it Planck} satellite \citep{planck13_p17}.  The larger {\it Planck} beam and higher map noise level (at 143 GHz, $\sim$ 7 arcmin and 45 $\mu$K-arcmin) lead to the much higher $5\sigma $ flux threshold of 145\,mJy at 143 GHz.  The larger beam corresponds to a much lower effective maximum multipole used in the lensing reconstruction ($\lmax \simeq 1800$), reducing the relative impact of foregrounds.  However, the higher flux cut increases the bias substantially.  The {\it Planck} team thus found a bias of amplitude $\sim 2\%$  on the lensing power spectrum from unresolved bright sources whose distribution approaches Poisson.  Since this corresponds to one-half of their 4\% statistical error, this term needed to be treated explicitly in their analysis.  This was done by estimating the amplitude for the trispectrum of these roughly uncorrelated sources directly from the maps, evaluating the associated bias on the lensing power spectrum using the curved-sky generalization of Eq.~\ref{eq:trisp_biascalc}, and subtracting the resulting template from the estimated lensing power spectra.

We note, however, that infrared galaxies do not need to show a Poisson-distributed component given the maximum multipoles measured by current experiments.  If measurements are made assuming they contain two separate components, clustered and Poisson, then there is a possibility that the amplitude of the Poisson component will be a function of the maximum multipole of the measurement (as is the case for the S10 model).  We discuss more realistic, clustered sources below.

\subsubsection{Clustered Galaxies}
\label{sec:clusteredgals}
The distribution of sources on the sky 
exhibits clustering, which affects the higher-order statistics.  Clustering in the CIB at mm-wave frequencies was first detected in the CIB power spectrum  by \citet{hall10}, \citet{dunkley11} and \citet{planck11_p18}, and has since been measured with increasing precision \citep{shirokoff11,reichardt12, sievers13,planck13_p30}.
The clustering in the bispectrum at mm-wave
frequencies has also been recently detected \citep{crawford13, planck13_p30}.  Although some
analytic prescriptions exist for  higher-order moments of
the galaxy population (e.g., \citealt{argueso03, lacasa12}), we use the 
S10 and B13  simulations to estimate lensing
biases from clustered CIB sources.

To study the impact of source masking, we mask pixels above different maximal flux values ranging between 0.5 and 20\,mJy, replacing masked sources with the median of the map at each step.  Although the B13 simulations do not contain individual sources, we nevertheless perform masking on the brightest pixels, using the same conversion of temperature to flux density per pixel as that applied for the S10 simulations.  We  downsample the S10 simulations from Healpix resolution $N_{\rm side} = 8192$ to $N_{\rm side} = 4096$; the B13 simulations are  at resolution $N_{\rm side} = 4096$ natively.  We then extract 42 non-overlapping flat-sky maps of 100 square degrees each  using the oblique Lambert
equal-area azimuthal projection \citep{snyder87} from these $N_{\rm side} = 4096$ maps.  To finely sample the Healpix maps on the flat-sky grid we use a resolution of 0.25\arcmin , and then downgrad to  1\arcmin, which resolution is set by that needed for lensing reconstruction, by averaging contiguous pixels.  We account for the effects of this downsampling process by generating Healpix maps of simulated white noise (with power spectrum $C_l \equiv 1$), passing these maps through the same downsampling and reprojection steps, and estimating the power spectra of the resulting maps.  This leads to an estimate of the effective transfer function associated with the downsampling, which we deconvolve from our flat-sky CIB maps.

The top-left panel of  Fig.~\ref{fig:cib_forpaper} shows the power spectra, estimated on the flat sky, of both sets of CIB simulations when applying a  5\,mJy flux cut.   The model used to generate the B13 simulations assume infrared galaxies have both a Poisson and clustered component, and only describes the clustered piece.   This leads to a lower overall power spectrum amplitude at $l = 3000$ compared to the S10 simulations.  Also shown at $l = 3000$ are the measurements of the clustered and total CIB power found by  \citet{reichardt12} and \citet{sievers13}, the latter of which is a fit to the power spectra found by  \citet{das13}.\footnote{The uncertainty  shown for the ``obs. total'' CIB points is obtained by combining in quadrature the uncertainties on the measured amplitudes of CIB sources fit to a model of infrared galaxies with separate clustered and Poisson components.  We neglect the  covariance between these measured amplitudes, slightly overestimating  the uncertainty on the total.}  In the case of the \citet{reichardt12} measurement, we also show an approximate systematic error bar associated with the scatter between the five  models of the clustered CIB studied in that work.

We perform lensing reconstructions on the CIB fields using the estimator in Eq.~\ref{eq:reconst_defl}.  The power spectra of the reconstructed fields are shown in the bottom-left panel of Fig.~\ref{fig:cib_forpaper}.   In each case, we subtract an estimation of the noise bias $\nlzero$ using Eq.~\ref{eq:wrongnormal} and the power spectrum $(C_l^t)^\prime$ estimated directly from the simulations.  We call the remaining bias the ``CIB 4-pt. bias'' due to its dependence on the four-point function of the CIB.  

\begin{figure}
  \includegraphics[width=3.5in]{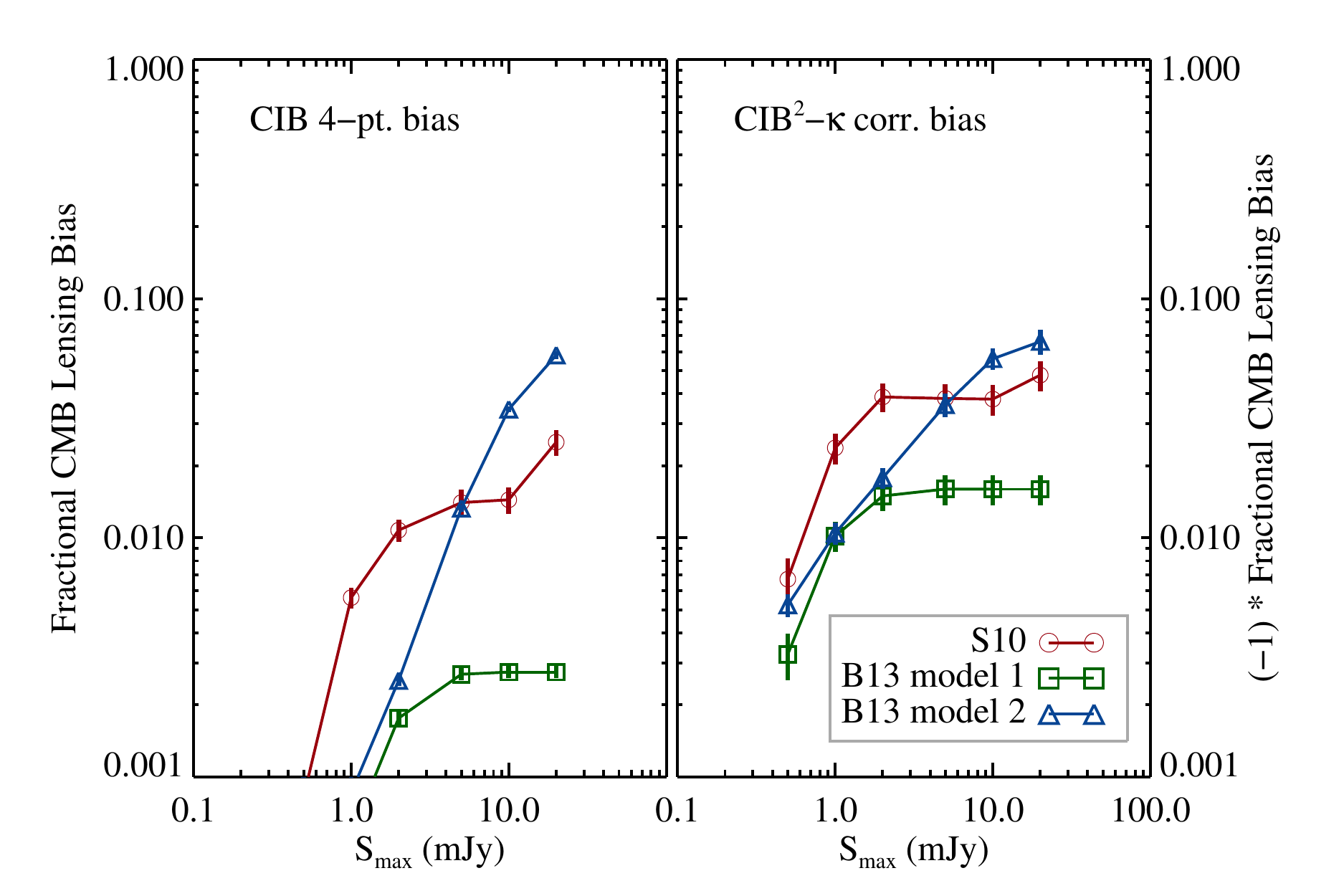}
  \caption{Fractional biases, in units of the lensing power spectrum,  evaluated at $L=500$ for the CIB models we consider.  Left panel: bias from the CIB four-point function; right panel: absolute value of the bias from the CIB-$\kappa$ correlation, which is negative on the scales of interest.  Error bars denote the error on the mean, based on the scatter from 42 patches, each of 100 square degrees.}
  \label{fig:cib_fractional}
\end{figure}

We show this bias, evaluated at $L=500$,   as  a function of flux threshold in  the left panel of Fig.~\ref{fig:cib_fractional}.   While for the unmasked fields the biases can be several percent, masking to $\sim 2$~mJy can reduce them to below one percent for all three simulations tested. Since a 5$\sigma$ flux cut at 150 GHz corresponds to $\sim 5$~mJy for an ACT/SPT experiment, it may be necessary to use information from higher frequencies to reliably extract sources to this level.

\subsection{The Galaxy-Lensing Correlation}
As was shown observationally by \citet{holder13}, \citet{planck13_p18}, and \citet{hanson13}, the CIB is strongly correlated with the CMB lensing field, at the $\sim 80\%$ level \citep{song03}.  This is because both fields are tracers of nearly the same dark matter fluctuations, owing to the very similar redshift responses.  The two  sets of simulations that we use include maps of the lensing convergence, $\kappa$, which is related to the lensing potential according to $\kappa(\nhat) = {1\over 2} \nabla^2 \phi(\nhat)$.  The $\kappa$ field is proportional to the projected matter overdensity along the line of sight.   In the top-right panel of Fig.~\ref{fig:cib_forpaper}, we show the $\kappa$-CIB cross-power spectra in the simulations.  Also shown is the cross-correlation coefficient of $\simeq 0.8$ at 143\,GHz found by the {\it Planck} team \citep{planck13_p18}.  We estimate the uncertainty on this cross-correlation amplitude  by summing, in quadrature, the inverse of the stated fractional bandpower uncertainties taken from Table~2 of \citet{planck13_p18}.  The inverse of the square root of this sum gives a fractional uncertainty of 22\%.\footnote{Note that this uncertainty is much smaller at the higher \Planck~frequencies, where emission from dust dominates over the CMB.}     The cross power spectra in the simulations are lower than those found observationally, since dark matter halos are only identified up to a given redshift in the simulations ($z=4$ for B13, $z=3$ for S10).  As a result, CIB sources do not populate halos at higher redshifts than these in the simulations.

We find that CIB fields that are correlated with the lensing maps lead to a bias on the lensing power spectrum \citep{smith07,vanengelen12,bleem12b}.  While trispectra based on these correlations, assuming the CIB is a Gaussian field, have been derived \citep{cooray03}, analysis of non-Gaussian CIB simulations has indicated that a much larger bias can exist.  
This bias is negative on the scales of interest ($L<2000$), and becomes positive at higher $L$.  To isolate this effect, we compute the cross power of the lensing reconstructions of CIB fields with the input $\kappa$ maps.  This bias is a multiplicative bias, describing the imperfect $\kappa$ reconstruction using maps with correlated non-Gaussian foregrounds.  This is in contrast to the additive bias of intrinsic four-point functions from foregrounds (as discussed in the previous section). We multiply our result by a factor of 2, since we are considering biases for lensing auto-power spectra.  This factor of 2 would be absent in an analysis of cross-correlations of CMB lensing with other, independent tracers of large-scale structure.  Since this bias is proportional to two factors of the CIB and one factor of $\kappa$, we call this the ``CIB$^2$-$\kappa$ bias.''  This bias probes the matter bispectrum as traced by these two fields.

The results from the simulations are shown, in absolute value, in the bottom-right panel of Fig.~\ref{fig:cib_forpaper}.  Here again there is substantial variation between the different simulations, with the bias factors ranging from 1\% to 5\%, as seen fractionally in the right panel of Fig.~\ref{fig:cib_fractional}.  This bias does not fall as quickly with masking as the intrinsic four-point bias, and exceeds $1\%$ even for aggressive
 thresholds of 1~mJy.

\section{Bias on Lensing Reconstruction from Galaxy Clusters}

In addition to infrared and radio galaxies, galaxy clusters also have an intrinsic four-point function and a correlation with the $\kappa$ field.  Both of these also lead to biases on the lensing power spectrum.

\subsection{The Galaxy Cluster Four-point Function}
Galaxy clusters are apparent in CMB maps due to the Sunyaev-Zel'dovich (SZ) effect. This effect is a spectral distortion in the CMB caused by the inverse Compton scattering of CMB photons with free electrons (\citealt{sunyaev70, sunyaev72}; for reviews see \citealt{rephaeli95, carlstrom02}).  In this work, we consider the thermal Sunyaev-Zel'dovich effect (tSZ), which involves photons scattering off hot electrons in the deep potential wells of galaxy clusters.  The kinetic Sunyaev-Zel'dovich effect (kSZ), which originates from photons scattering off electrons possessing bulk motions along the line of sight, is expected to be  a much smaller contaminant for CMB lensing studies \citep{vanengelen12}.  This is due to its  lower   fluctuation amplitude and its smoothness compared to the tSZ effect.  In addition, masking of tSZ clusters (discussed below) will reduce the kSZ-induced lensing contamination from those clusters \citep{amblard04}.

The non-Gaussianity of the tSZ field is significant; the three-point
function, or bispectrum, has recently been detected by the ACT
\citep{wilson12}, SPT \citep{crawford13}, and {\it Planck}
\citep{planck13_p21} teams.  For lensing power spectrum estimation the tSZ
trispectrum is an important potential contaminant. The tSZ trispectrum has been 
considered previously as a source of non-Gaussian variance in CMB
power spectrum estimation \citep{cooray01b, shaw09}.  Due to its
scaling with the fourth power of the temperature decrement, it is
dominated by the most massive clusters in the Universe.  In the
current era of dedicated SZ surveys with low noise levels, namely SPT
\citep{vanderlinde10,benson13} and ACT \citep{
  sehgal11,hasslefield13}, many of these massive clusters can be
detected on an individual basis.  However, the trispectrum from
clusters just below the detection threshold can in principle cause a
concern for lensing studies.

The power spectrum of the tSZ effect has been extensively modeled
\citep[e.g.,][]{komatsu02, sehgal10, shaw10, battaglia10}, and observations of the
power spectrum of the mm-wave sky have been used to fit these
models \citep[most recently,][]{ reichardt11,sievers13,planck13_p21}.
The power spectrum is dominated by clusters with lower masses than
those which can be detected individually in current surveys \citep{shaw09,trac11}.  The bispectrum of the tSZ effect originates from clusters with masses between those dominating the power spectrum and trispectrum, and can provide a useful cosmic probe \citep{bhattacharya12,hill13}.

Here, we compute the impact of the tSZ trispectrum on
lensing power spectrum estimation using the same halo model approach as that taken in
the predictions for the SZ power spectrum and bispectrum.  On the scales
of interest, the power spectrum and bispectrum have been found to be dominated by the term in which all multipole arguments reside within
the same dark matter halo \citep{komatsu99a, bhattacharya12}; this is known as the ``one-halo'' term.  We
expect the one-halo term to dominate here as well, due to the
sensitivity of the trispectrum to the brightest objects

The tSZ trispectrum is then given by \citep{cooray01b, komatsu02, bhattacharya12}
\begin{align}
  \trisp(\lvec_1, \lvec_2, \lvec_3, \lvec_4) = & f_\nu^4 \int dz {dV \over dz} \int d \ln M {dn(M,z) \over d \ln M} \nonumber \\ & \times y(\ell_1, M, z) y(\ell_2, M, z) y(\ell_3, M, z) y(\ell_4, M, z). \label{eq:sztrisp}
\end{align}
Here, $V(z)$ is the comoving volume element per steradian, $f_\nu = x_\nu (\coth(x_\nu) - 4)$ is the frequency-dependent SZ scaling factor, $x_\nu = h \nu / k_B T_{CMB}$, $n(M, z)$ is the halo mass function, which we take to be that of \citet{tinker08}, and $y(\ell, M, z)$ is the 
Fourier transform of the projected SZ profile for a cluster mass $M$ and redshift $z$.  This last term is given  by \citep[e.g.][]{komatsu02}
\begin{equation}
  y(l, M, z) = {4 \pi r_s \over l_s^2} {\sigma_T  \over m_ec^2} \int dx x^2 P_e(x, M, z) {\sin(lx/l_s) \over lx / l_s}.  
\end{equation}
Here, $r_s $ is the scale radius of the pressure profile, $l_s = D_A(z)
/ r_s$ where $D_A(z)$ the angular diameter distance to redshift $z$,
$\sigma_T$ is the cross-section for Thomson scattering, $m_e c^2$ is
the electron rest mass, $x = r / r_s$, and $P_e(x, M, z)$ is the projected pressure profile of
the cluster, for which we use the profile described in \citet{bhattacharya12}.

\begin{figure}
  \includegraphics[width=3.5in]{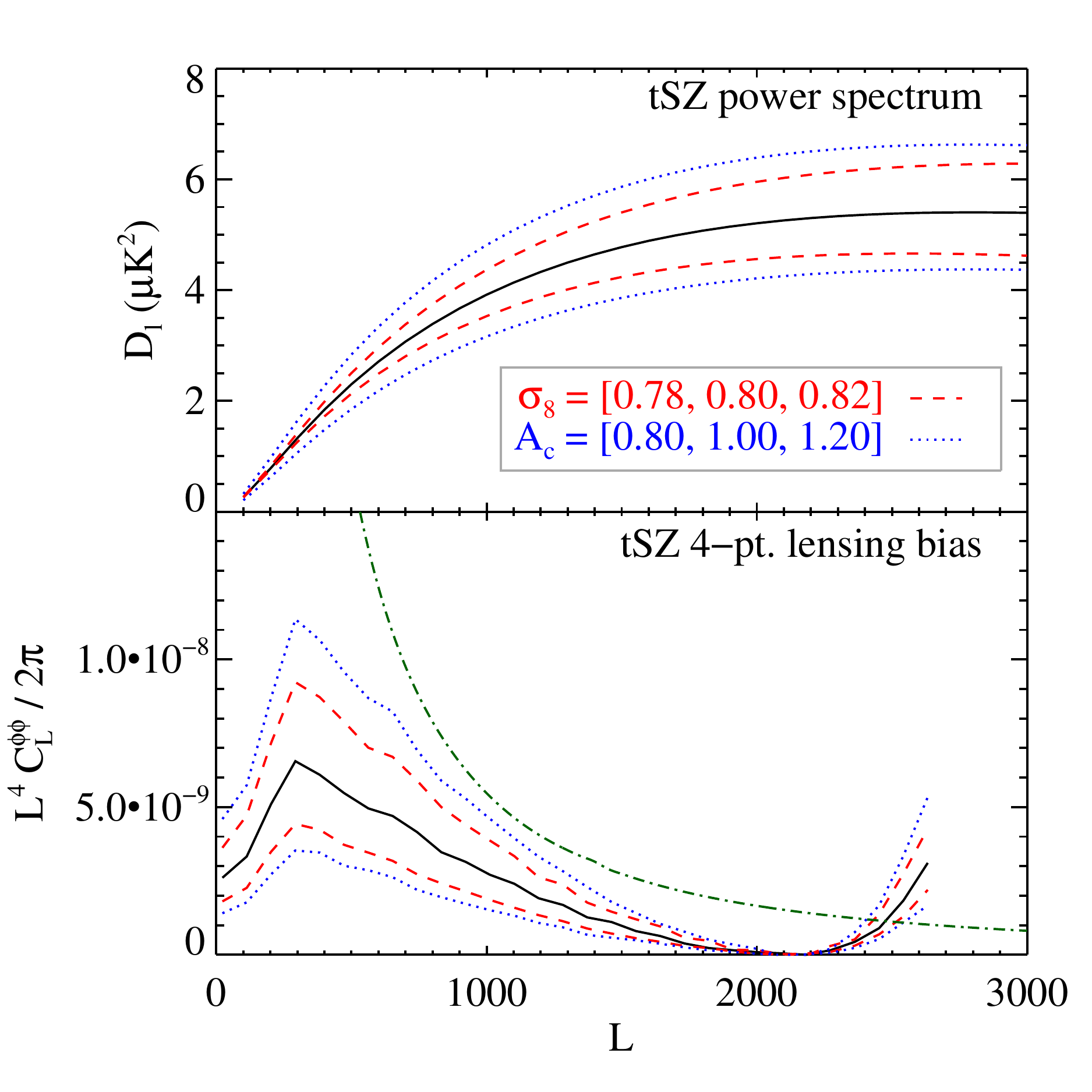}
  \caption{ Impact on tSZ power spectrum (top) and tSZ four-point lensing bias (bottom) from varying the parameters describing the amplitude of fluctuations $\sigma_8$ (red, dashed) and the normalization of the concentration-mass relation $A_c$ (blue, dotted).}
  \label{fig:cm_s8_forpaper}
\end{figure}

To evaluate the bias on the lensing estimator, we insert the SZ
trispectrum into Eq.~\ref{eq:trisp_biascalc}. We consider halo masses in the range $2\e{13} \Msol
\leq \mvir \leq 5\e{15}\Msol$, and redshifts in the range $0<z<3$.
The resulting features in the lensing power spectrum space, shown in Fig.~\ref{fig:cm_s8_forpaper}, have very similar shapes as those for Poisson-distributed point sources, but the spatially extended nature of clusters reduces the amplitude of this template.  In Fig.~\ref{fig:cm_s8_forpaper}, we also show the theoretical thermal SZ power spectrum, computed using the analogue of Eq.~\ref{eq:sztrisp} with two factors of $y(l,M,z)$ (e.g.,~\citealt{komatsu02}). 

This analytic approach enables us to evaluate the SZ four-point bias as a function of both the  cluster physics parameters and the  cosmological parameters.  Combinations of these parameters are constrained by measurements of the tSZ power spectrum \citep{reichardt11,dunkley11,planck13_p21}.  Perturbing the cluster physics parameters described in \citet{bhattacharya12} and evaluating the lensing bias, we find that the  parameter with the largest impact  is the normalization of the concentration-mass relation, $\ac$.  This parameter controls the effective radius of the clusters.  The amplitude of the matter power spectrum today, $\sate$, also has a strong impact on the moments of the tSZ field, with the power spectrum scaling in proportion to $\sate^{7-9}$, and the bispectrum scaling in proportion to $\sate^{11-12}$ \citep{bhattacharya12,hill13}.  In Fig.~\ref{fig:cm_s8_forpaper} we show the dependence of  the power spectrum and the four point-induced lensing bias on both of these parameters.  As expected, the fractional changes in the four-point signature are roughly  twice those of the power spectrum, leading to a large uncertainty.

Since variation in these parameters can lead to a significant  change in the theoretical tSZ four point-induced lensing bias, we compute this feature on a grid of $10 \times 10$ points in the plane formed by $\sate$ and $\ac$.  We assume a range of parameter values of $0.70 < \sate < 0.85$ and $0.3 < \ac < 2.0$, and impose prior information from measurements of the tSZ power spectrum on the angular scales of interest.  Specifically, we use the SPT measure from \citet{reichardt11} of $D_{3000}^{\rm tSZ} = 3.65 \pm 0.69\,\mu$K$^2$, where $D_l \equiv l(l+1)C_l/2\pi$.  This measurement is obtained from fitting the amplitudes of multiple  templates to power spectra at three frequencies, and is somewhat sensitive to choices of these templates, particularly the properties of the CIB.  To be conservative we have chosen the tSZ template fit amplitude from \citet{reichardt11} with the widest error bar.  We form a simple chi-square-like function defined by 
\begin{equation}
  \chi^2(\sate, \ac) = \left ( C_{3000}^{\rm tSZ,data} -  {C_{3000}^{\rm tSZ,theory}(\sate,\ac)  } \over \sigma(C_{3000}^{\rm tSZ,data}) \right )^2.
\end{equation}
The theory power spectra, $C_{l}^{\rm tSZ,theory}$, are calculated without any clusters masked, i.e.,~over all clusters in the range of $2\e{13}\,\msol \leq \mvir \leq 5\e{15}\,\msol$.   We find that this function has a minimum along a long degeneracy track in the $\sate-\ac$ plane, given by the equation $\ac - 1.0 = -13 \times (\sate - 0.76)$.  

For a variety of cluster mass thresholds, we compute the tSZ four point-induced lensing bias at each point in our grid of values of $\sate$  and $\ac$ (Eqs.~\ref{eq:sztrisp} and \ref{eq:trisp_biascalc}).  We then calculate the scatter among these models subject to a data-derived weighting function given by $w(\sate, \ac) = \exp( - \chi^2(\sate, \ac) / 2)$.  This leads to a data-constrained estimate of the theoretical model scatter, effectively marginalizing over these parameters.

\begin{figure}
  \includegraphics[width=3.2in]{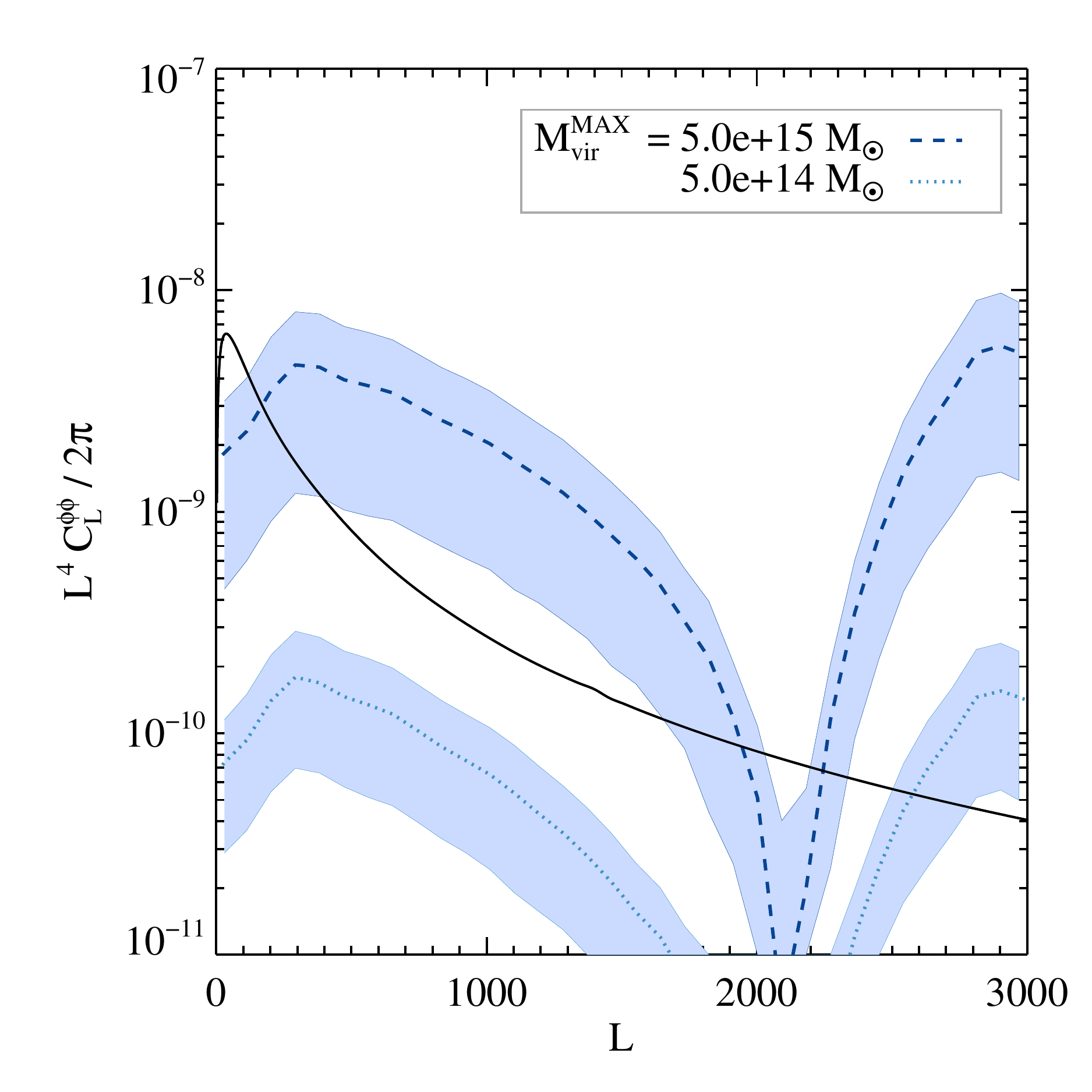}
  \caption{ Bias from the four-point function of the thermal SZ effect, obtained theoretically for two mass thresholds.  The error bands indicate the impact of marginalizing over $\sigma_8$ and $A_c$, while forcing the theoretical power spectrum to agree with measurements from SPT \citep{reichardt11}, within uncertainties.}
  \label{fig:tsz_theory_witherrors}
\end{figure}
\begin{figure}
  \includegraphics[width=3.5in]{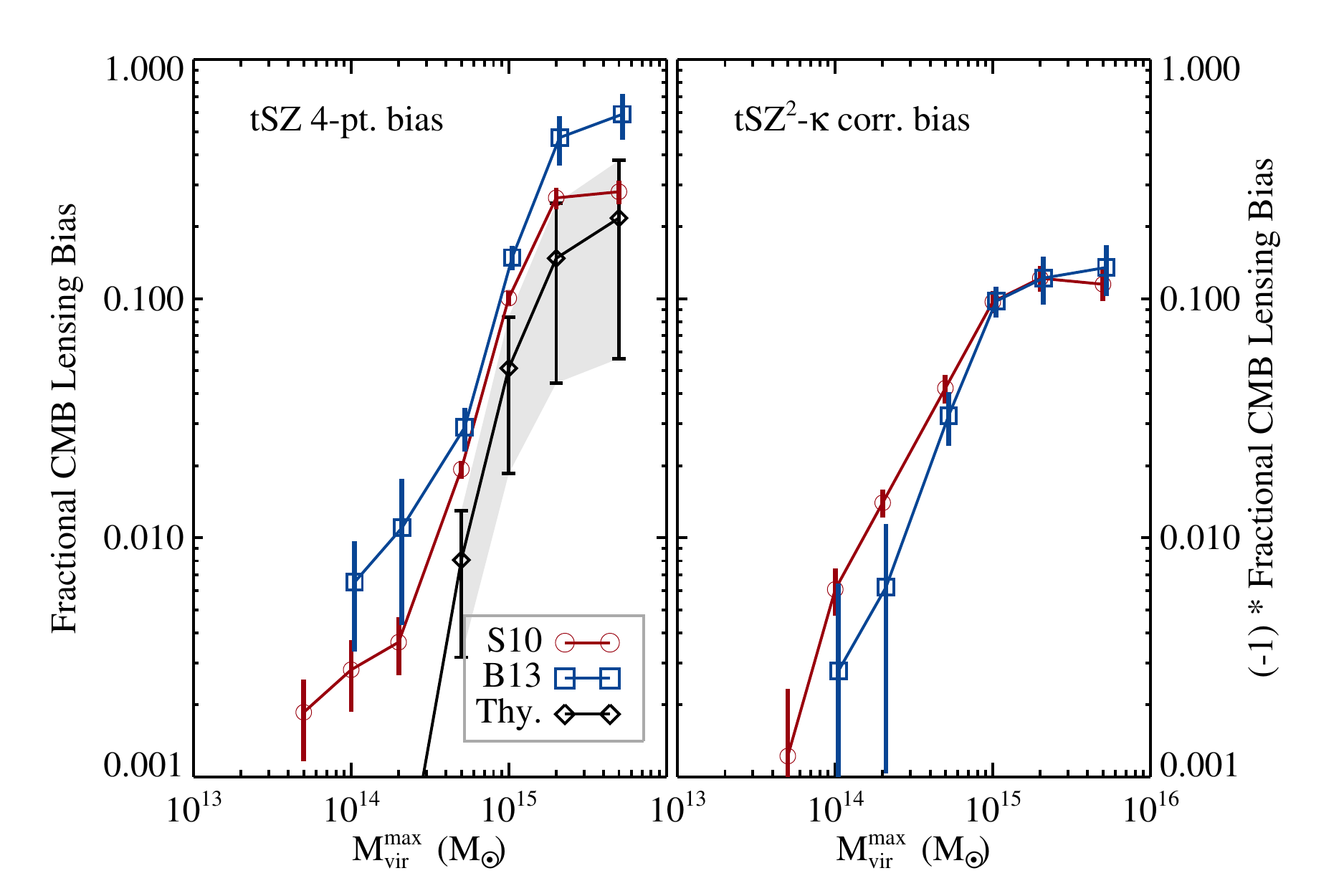}
  \caption{Amplitude of simulation-derived thermal SZ   biases at $L=500$.  Left panel: bias from tSZ four-point function; right panel: absolute value of the (negative) bias from the cross-correlation between tSZ and $\kappa$.  Error bars denote the error on the mean, based on the scatter from 42 patches, each of 100 square degrees.  In the left panel, the black points correspond to the theoretical biases from the tSZ trispectrum, Eqs.~\ref{eq:sztrisp} and \ref{eq:trisp_biascalc}, with errors that include marginalization over $\sate$ and $\ac$.  The amplitude is lower than in the simulations because the simulations use a higher $\sate$ and $\ac$ than the theory. }
  \label{fig:tsz_fractional}
\end{figure}

In Fig.~\ref{fig:tsz_theory_witherrors}, the resulting error bands are shown as a function of lensing multipole, $L$, for two choices of mass cut.  The values at $L=500$ are also shown, along with the simulation-derived biases described  below, in the left panel of Fig.~\ref{fig:tsz_fractional}.    The theory-derived biases on the CMB lensing power spectrum correspond to $(0.0013\pm 0.0004 \%, 0.80 \pm 0.49 \%, 21 \pm 16\%)$ for cluster mass thresholds of $\mvir = (10^{14}, 5\e{14}, 5\e{15})\msol$, respectively.

\begin{figure}
  \includegraphics[width=3.5in]{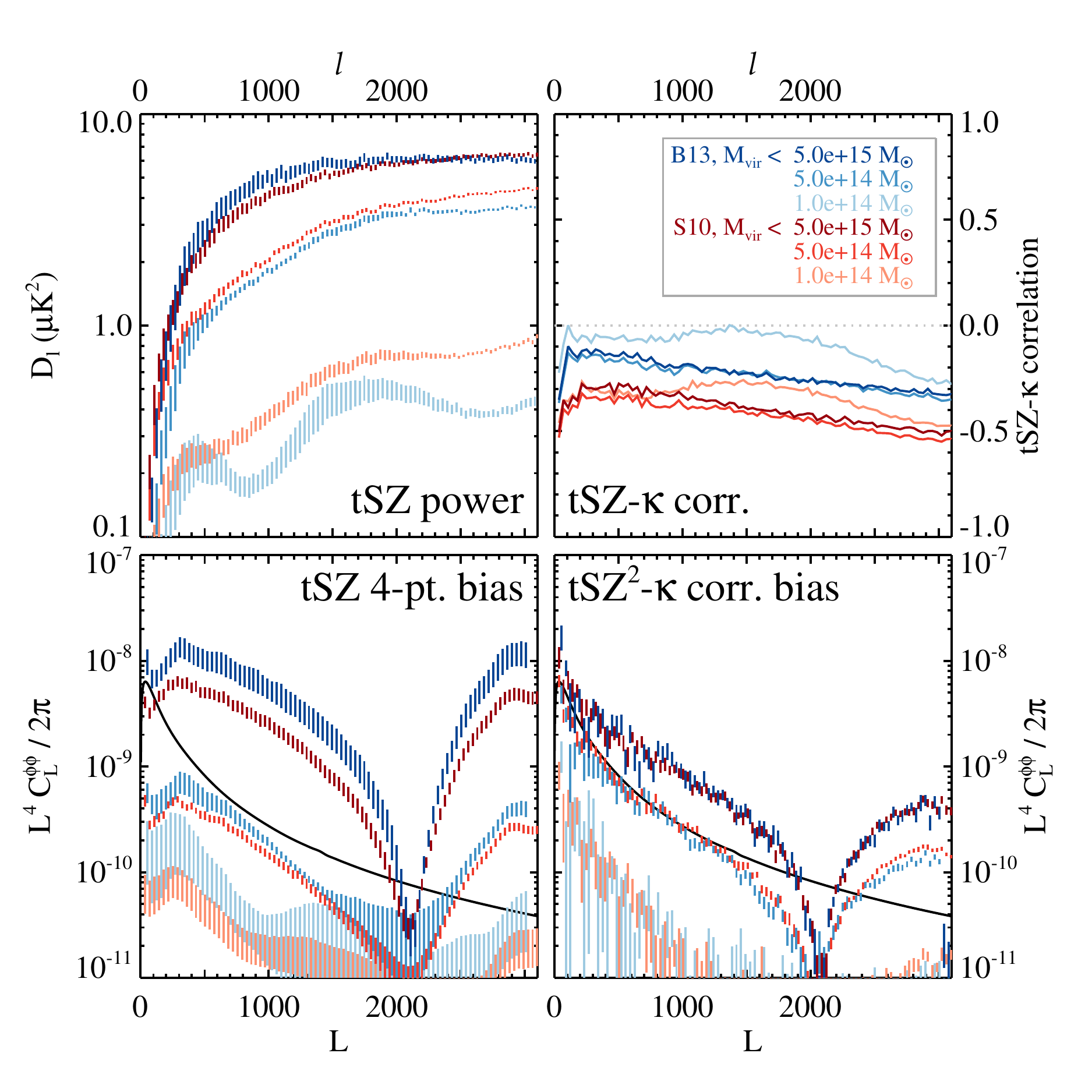}
  \caption{Biases on CMB lensing power spectra from tSZ clusters.  Top left: power spectra of the tSZ simulations, including those of B13 (blue) and S10 (red).  Fainter colors correspond to more aggressive cluster mass cuts.   Bottom left: CMB lensing bias from the four-point function of these reconstructions as a function of lensing multipole, $L$.   Error bars denote the error on the mean, based on the scatter from 42 patches of 100 square degrees each. Top right:  Cross correlation between the tSZ and the lensing field $\kappa$ in the simulations.  Bottom right: absolute value of the  bias induced from correlation between the square of the tSZ and the lensing field.  The black curves show the lensing power spectrum multiplied by 0.05.}
  \label{fig:tsz_forpaper}
\end{figure}

In addition to the analytical approach, we  use simulations to
estimate the tSZ four-point bias.  As described for the CIB in Section
\ref{sec:clusteredgals}, we project 42 patches, each of 100 square degrees, of
the tSZ Healpix maps from the S10 and B13 simulations onto the flat
sky, and perform lensing reconstruction.  We perform cluster masking by
setting pixels within 5' of the cluster center to the median of the pre-masked map.

Figure \ref{fig:tsz_forpaper} shows the tSZ power spectra and
four-point biases from these simulations.  The latter are obatined by
computing the power spectrum of the reconstructed lensing maps.  We
choose to mask using the virial mass for each cluster, since this mass
definition is better matched to our theoretical model.  While the
amplitudes of the power spectra of the two simulations are comparable,
the amplitude of the four-point feature differs by a factor of $\sim
1.6$.  This is commensurate with real-space measures of kurtosis  in the maps \citep{hill13}, which differ by a factor of 1.66.

The left panel of Fig.~\ref{fig:tsz_fractional} shows the fractional bias on the lensing power spectrum at $L=500$ for the two tSZ simulations.  In order for the four-point bias to be reduced to sub-precent levels we find that it is necessary to mask to $\mvir \lesssim 5\e{14} \Msol$.


\subsection{The Galaxy Cluster-Lensing Correlation}
As with the CIB, the thermal SZ sky is expected to be correlated with the CMB lensing field.  Since both sets of large-scale structure-based simulations contain lensing fields which are obtained from the same dark matter that is used to populate the halos, both will contain a nonzero cross-correlation.  As shown in the top-right panel of Fig.~\ref{fig:tsz_forpaper}, the S10 simulations (with the updated tSZ model  described in Section~\ref{sec:simulations}) yield a tsz-$\kappa$ cross-correlation coefficient  of $40\%$, and the B13 model yields a cross-correlation coefficient of $20\%$.   These values are for effectively no cluster masking; the reduction in the cross-correlation when masking is performed is also shown.  The factor of two difference between the tSZ-$\kappa$ cross correlation obtained from the S10 and B13 simulations is likely due to differences in the modeling of the tSZ effect from the intergalactic medium and at high redshifts.Ê   

Performing lensing reconstructions on the tSZ fields and cross-correlating with the input lensing maps leads to a bias which is negative at $L<2000$, and positive at higher $L$, as with the CIB.  In the bottom-right panel of Fig.~\ref{fig:tsz_forpaper}, we show the bias from this correlation, for the two sets of simulations, and for three levels of cluster masking.  The values of the bias at $L=500$ are also shown as a function of mask level in the right panel of Fig.~\ref{fig:tsz_fractional}.  Without cluster masking, this is a $\sim10\%$ bias for both simulations.  Masking to $\mvir = 5\e{14} \Msol$ reduces the bias to $\sim 2\%$, and more aggressive masking reduces this bias to a sub-percent level.  


\subsection{Dependence of SZ Bias on Mask Radius}
\begin{figure}
  \includegraphics[width=3.5in]{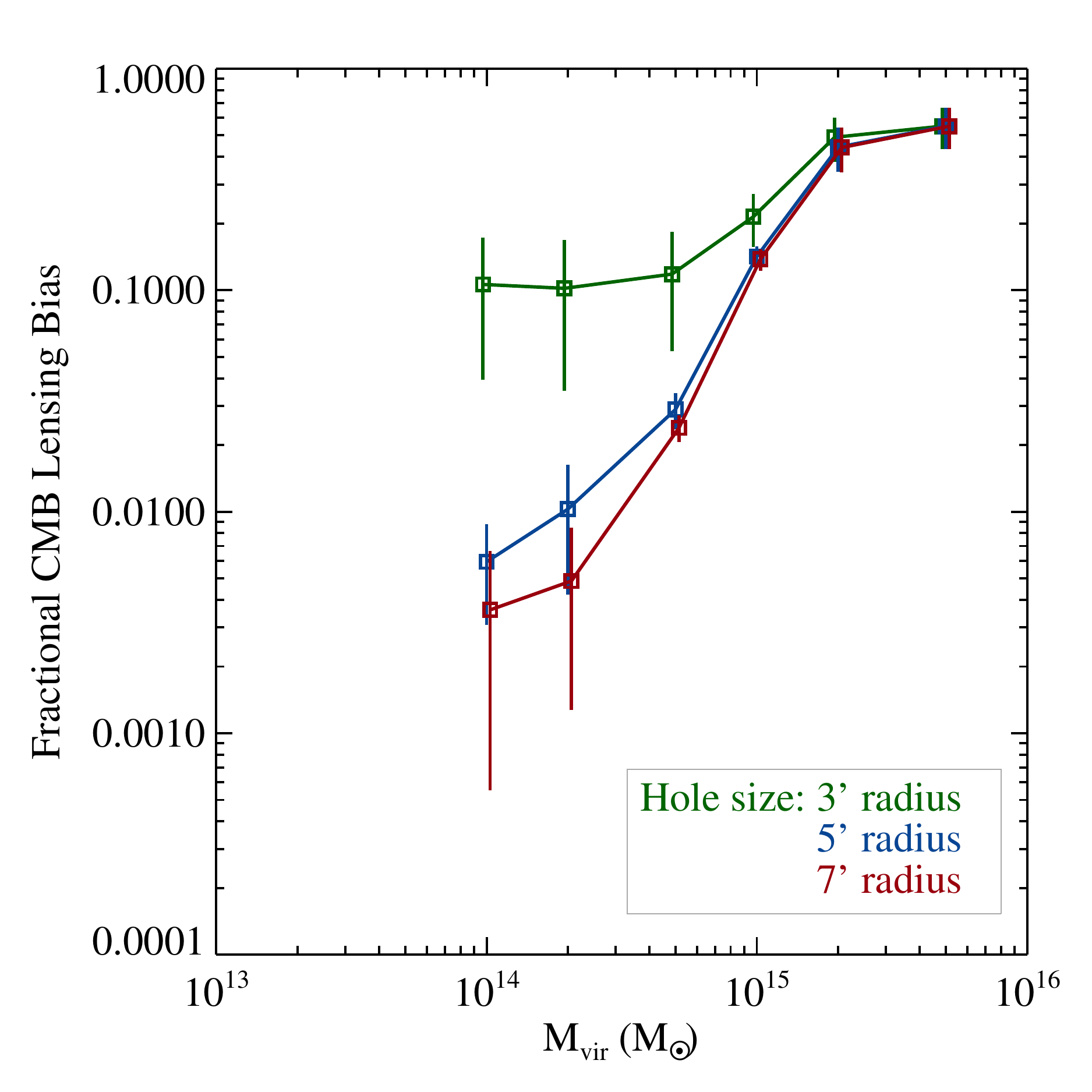}
  \caption{ Thermal SZ four-point lensing bias at $L=500$ in the B13 simulations for three cluster mask radii. }
  \label{fig:radii}
\end{figure}

Much of the tSZ four-point bias originates from large, relatively nearby halos.  An insufficiently large mask leaves wings around each large projected cluster, which can become the dominant source of non-lensing fluctuation in reconstructed lensing maps.  In Fig.~\ref{fig:radii}, we show the bias on the lensing power spectrum at $L=500$ as a function of  maximal cluster mass, for various mask sizes.  The plateaus seen at low mass correspond to incomplete masking of large halos, and it is clear that a mask of at least 5' radius is necessary for massive clusters, for percent-level accuracy on the lensing power.  Thus multiple mask sizes may be needed, with larger masks for nearby clusters.

\section{Dependence of lensing biases on maximal temperature multipole}
\label{sec:lmax}
The optimal filter for isolating lensing effects in quadratic CMB lensing reconstruction, Eq.~\ref{eq:filter}, naturally downweights the modes in the observed sky with the largest variance.  This can be seen by the presence in the denominator of the filter of the total power spectrum, $C_l^t$, which consists of the CMB, foreground, and noise power.  Treating the CMB as being beam-deconvolved, the noise power spectrum $C_l^{\rm noise}$ increases exponentially at  the beam scale, leading to a natural cutoff for the modes included in the lens reconstruction.  However, for experiments with high angular resolution this weighting can introduce the effects of foreground trispectra which become large at high multipoles, $l\gtrsim 2000$, where the CMB becomes relatively faint due to diffusion damping.  For the analysis in all preceding sections we have thus imposed a multipole limit of $\lmax = 3000$.  In this section we study  the dependence of the foreground biases on this choice.

\begin{figure}[h]
  \includegraphics[width=3.5in]{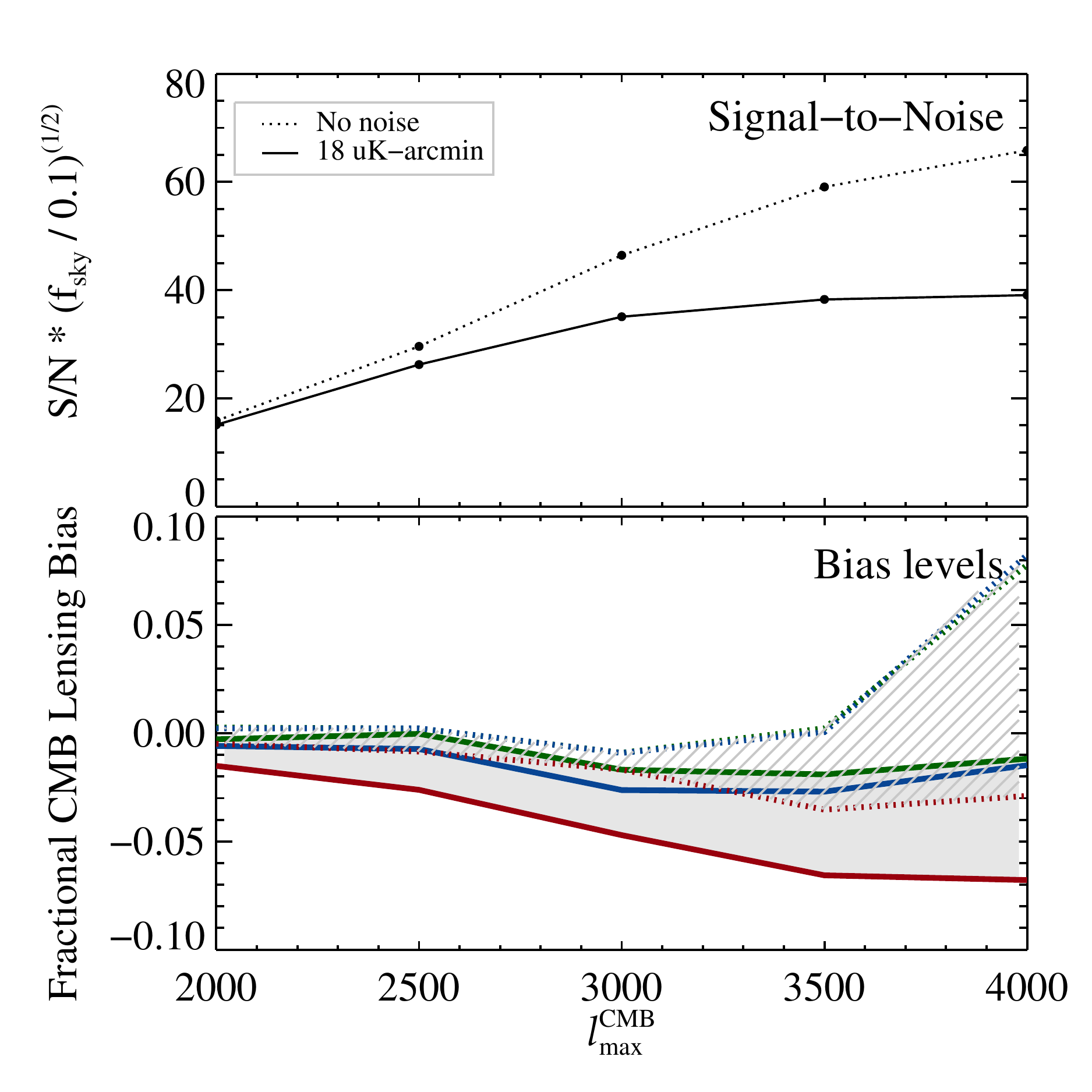}
  \caption{Statistical errors and bias levels as a function of maximum multipole included in the CMB lensing reconstruction.  Top panel: projected CMB lensing detection significance for an experiment with a noise level of 18\,$\mu$K-arcmin (solid) and a noise-free experiment (dotted).  In both cases, there is an additional effective white noise level of $9.1\,\mu$K-arcmin from the assumed level of foreground power. Bottom panel: The solid grey shaded region corresponds to the range of total bias expected for the 18\,$\mu$K-arcmin experiment, and the three thick lines correspond to the total biases for each of the three simulations we consider:  the S10 simulations (red),  the B13 simulations using CIB model 1 (green),  and the B13 simulations using CIB model 2 (blue).  The hatched grey region shows the total biases for the aggressive masking level of 1\,mJy and $10^{14}\msol$, with no instrumental noise.  The dotted curves correspond to the total biases for each of the three simulations.
}
  \label{fig:versus_lmax}
\end{figure}

In Fig.~\ref{fig:versus_lmax},, we show the statistical detection significance for quadratic lensing reconstruction when assuming only statistical errors, neglecting foregrounds, for both an experiment with white noise of amplitude  $18 \mu$K-arcmin noise and a 1' beam, and a no-noise experiment.  The statistical errors are calculated following
\begin{equation}
  L^2 (\Delta (\clphi))^2 = {1 \over {L \Delta_L f_{\rm sky}}}(L^2\clphi + \nlzero(\lmax)), 
\end{equation}
where $\Delta_L$ is the binning size and $f_{\rm sky}$ the fraction of sky observed.  The  statistical significance for a lensing detection is then estimated using  $ {S / N} = (\sum_L {  (\clphi / \Delta(\clphi))^2})^{1/2}$.  For the 18\,$\mu$K-arcmin experiment,  increasing the maximum multipole from $\lmax = 3000$ to $4000$ leads to an increase in signal-to-noise ratio of only 11\%, while for the no-noise experiment the gain is a factor of 1.35.  We also show the total bias determined for our simulations.

\begin{figure}[h]
  \includegraphics[width=3.5in]{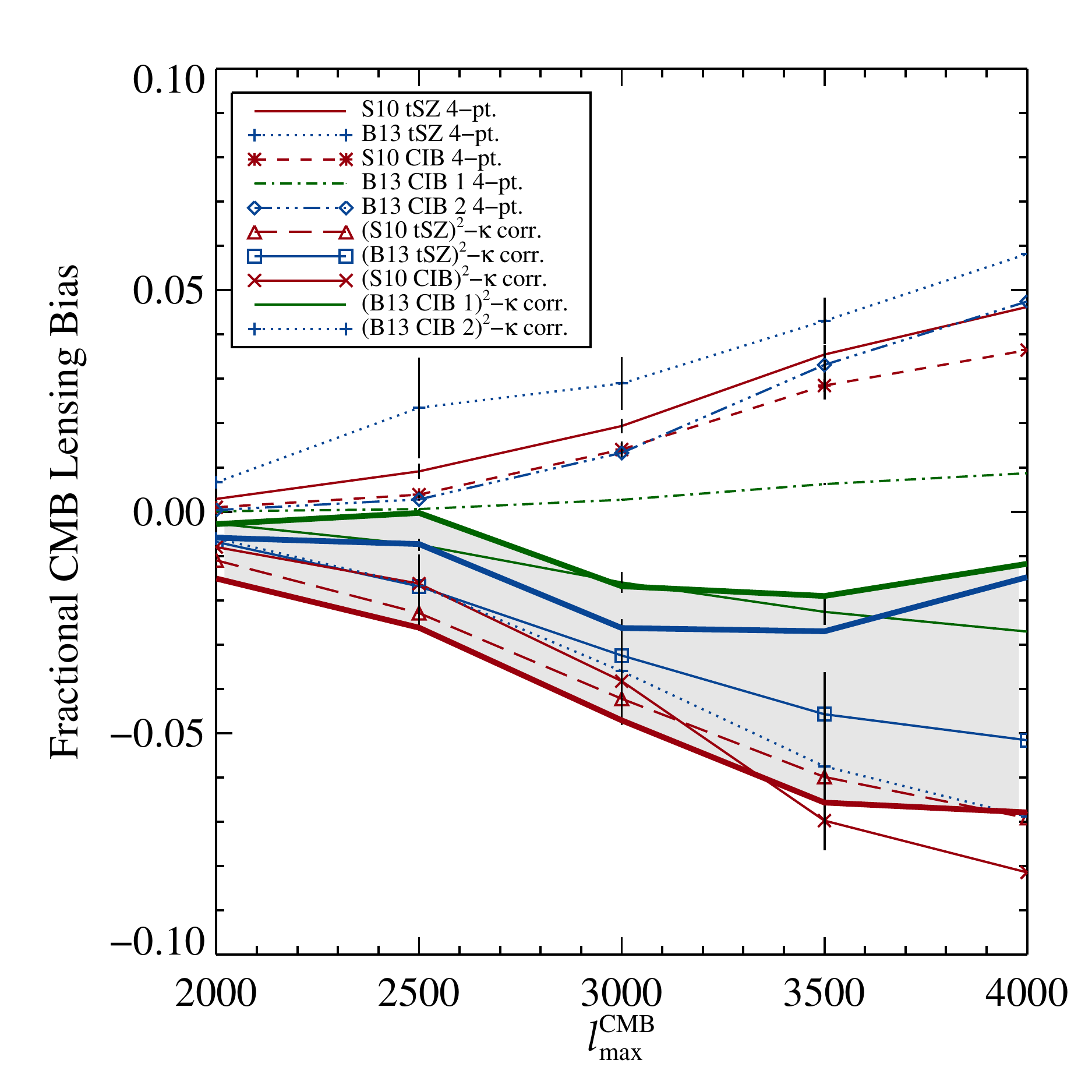}
  \caption{Simulation-based lensing reconstruction biases evaluated at $L=500$, as a function of the maximum multipole included in the temperature field, $\lmax$.  Curves shown are for an experiment with a noise level of 18\,$\mu$K-arcmin. The three thick lines, which delineate the grey shaded region, correspond to the total biases for each of the three simulations we consider:   the S10 simulations (red),  the B13 simulations using CIB model 1 (green),  and the B13 simulations using CIB model 2 (blue).  Also shown are the individual contributions to the total bias from the galaxy and cluster four-point functions and the galaxy and cluster correlations with the $\kappa$ field, for each simulation.
  }
  \label{fig:versus_lmax_v2}
\end{figure}

In Fig.~\ref{fig:versus_lmax_v2}, we show all the simulation-derived lensing biases as a function of $\lmax$, for an experiment with a noise level of 18\,$\mu$K-arcmin.  Here clusters are masked  to $\mvir = 5\e{14}\msol$ and sources masked to $5\,$mJy.  All biases can be seen to increase quickly with $\lmax$.  Also shown is the totals band (solid grey), bounded by the spread of the sum of the biases for each model.

\section{Reducing Bias by Aggressive Source Masking}
\label{sec:masking}




As we have shown, one way to reduce astrophysical biases on the lensing power spectrum is to mask sources and clusters aggressively, particularly, to lower detection thresholds than the very strongly-detected sources which would likely be masked in a standard analysis.  This process may introduce new biases, and in this section we use simulations of the lensed CMB to determine the response to aggressive masking on the estimation of the lensing power spectrum.

Several approaches have been put forward for dealing with source masks in practice.  One approach is to combine source masks with the mask from the edges of the field and any region of bright Galactic emission.  This can be treated as an additional source of statistical anisotropy, for which one can  apply the bias-hardened estimator of \citet{namikawa12}. With this technique, the effects of lensing and the mask multiplication are both treated as separate sources of statistical anisotropy, and optimized quadratic estmators for each are formulated.  An unbiased lensing estimate, which is valid to first order in the masking, can then be formed with a suitable linear combination of the two reconstructed fields.  

Another, more complete approach is to perform lensing reconstruction taking into account the full pixel covariances in the CMB maps.  Here the pixels which are to be masked can be assigned infinite variance, effectively projecting them out of the analysis.  For large CMB maps this approach is naively very computationally challenging, but can be sped up with appropriate preconditioning  \citep{smith07, smith09}.

A third method for treating sources is to restore the continuity of the CMB map at the source locations with in-painting techniques.  This  was studied by \citet{perotto10} and \citet{benoitlevy13}, using constrained Gaussian realizations to match the CMB fluctuations near the holes \citep{hoffman91}.  \citet{benoitlevy13} found that for masks of up to 2\% of the  sky, a straightforward rescaling of the \nlzero-subtracted reconstructions returns the input lensing power spectrum to good accuracy.  

\begin{figure}
  \includegraphics[width=3.5in]{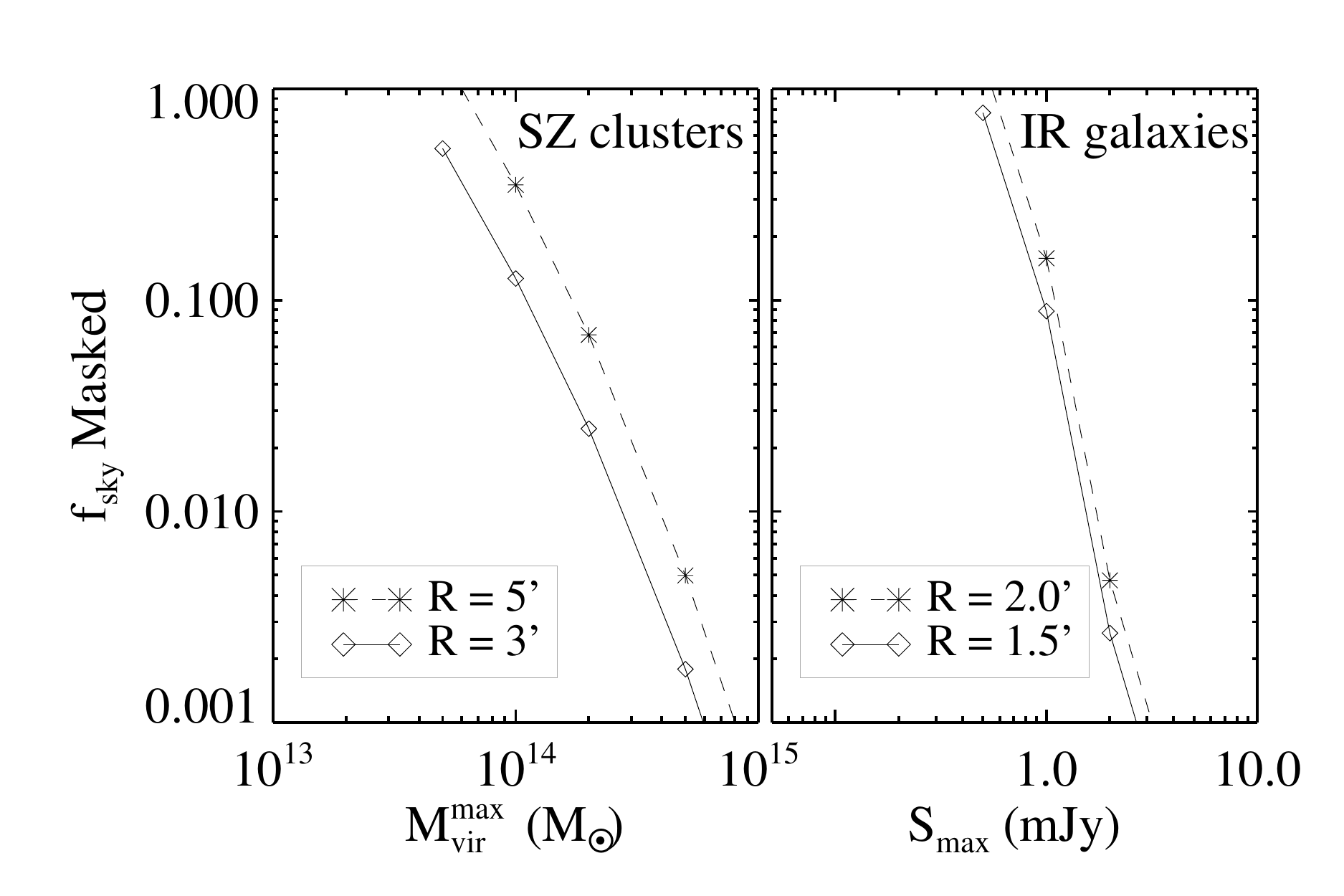}
  \caption{Fraction of sky masked in the S10 simulations, using the given hole radii, for different mass and flux thresholds.  Left: SZ clusters; right: Infrared galaxies.}
  \label{fig:fskymasked}
\end{figure}

As can be seen in Fig. \ref{fig:fskymasked},  the total area of sky masked increases rapidly with aggressive masking, such that, depending on the mask radius, masking sources to 1\,mJy and clusters to $\mvir = 10^{14} \msol$  corresponds to  10-30\% of the sky.

\begin{figure}
  \includegraphics[width=3.5in]{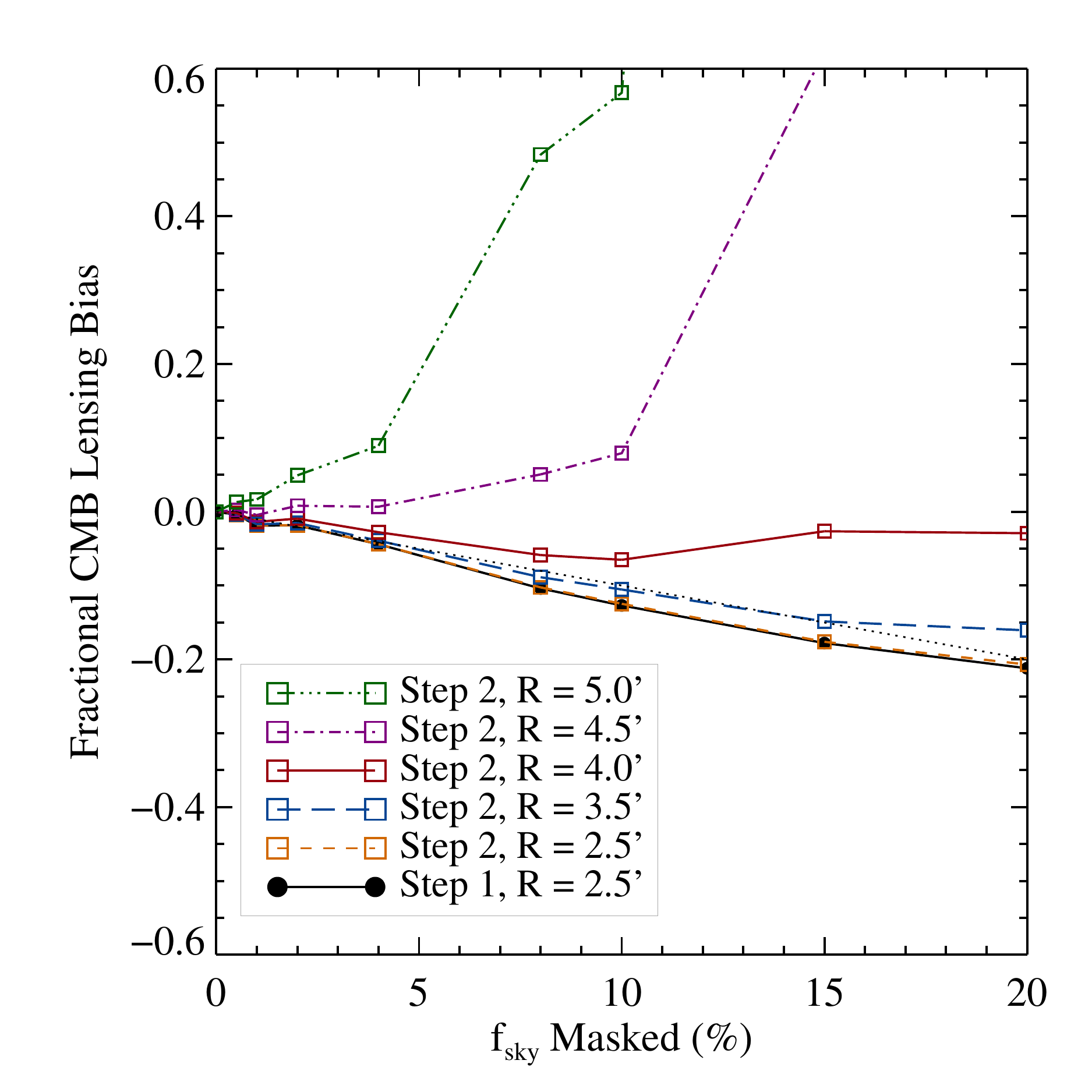}
  \caption{Fractional lensing biases at L=500 as a function of the fraction of sky masked when in-painting over masked regions with constrained Gaussian realizations.   The solid black line indicates the bias when only in-painting the CMB fields. Each colored line corresponds to additionally in-painting the reconstructed lensing deflection maps, at the same hole centers using hole sizes of different radii.
  }
  \label{fig:vs_fsky_masked}
\end{figure}

We perform a constrained realization of the structure in the masked regions using the routine described in \citet{bucher12}.  This routine takes as input the power spectrum of the sky, which is assumed known;  in a real analysis one would estimate the power spectrum directly from the valid data.  Performing lensing reconstruction and removing the \nlzero bias, we find that for holes of radius 2.5 arcmin and small fractions of the sky masked, the lensing power is reduced approximately in proportion to the amount of sky masked.  However, as seen in Fig. ~\ref{fig:vs_fsky_masked}, there is a significant deviation from this for some larger masked area fractions.


A straightforward approach for correcting for this bias would be to estimate it from Monte Carlo simulations.  We studied an alternative approach which can in principle use the data directly: we in-paint a second time, this time on the reconstructed lensing deflection maps.  Fig.~\ref{fig:vs_fsky_masked} shows this fractional biases at $L=500$.  In each case the maps of lensed CMB and noise are masked with holes of 2.5\arcmin~radius.  The reconstructed lensing fields are then in-painted again for a variety of mask sizes.
Since this second in-painting is restoring lensing power to the masked regions, one might expect that the bias should be significantly reduced.  However, we find that this is the case only when increasing the size of the holes to $\sim4\arcmin$~in this step.  Thus there is some significant nontrivial structure in the lensing maps in the vicinity outside each original point source mask.   Fig.~\ref{fig:vs_fsky_masked} shows that the residual bias can be greatly reduced using this two-step in-painting procedure, as in the case of using 4\arcmin~holes to in-paint the deflection map.  This leaves a smaller remaining bias that still needs to be evaluated using Monte Carlo simulations.  However, other techniques may still be devised to deal with this bias in a cleaner fashion.

\begin{figure*}
  \includegraphics[width=7.6in]{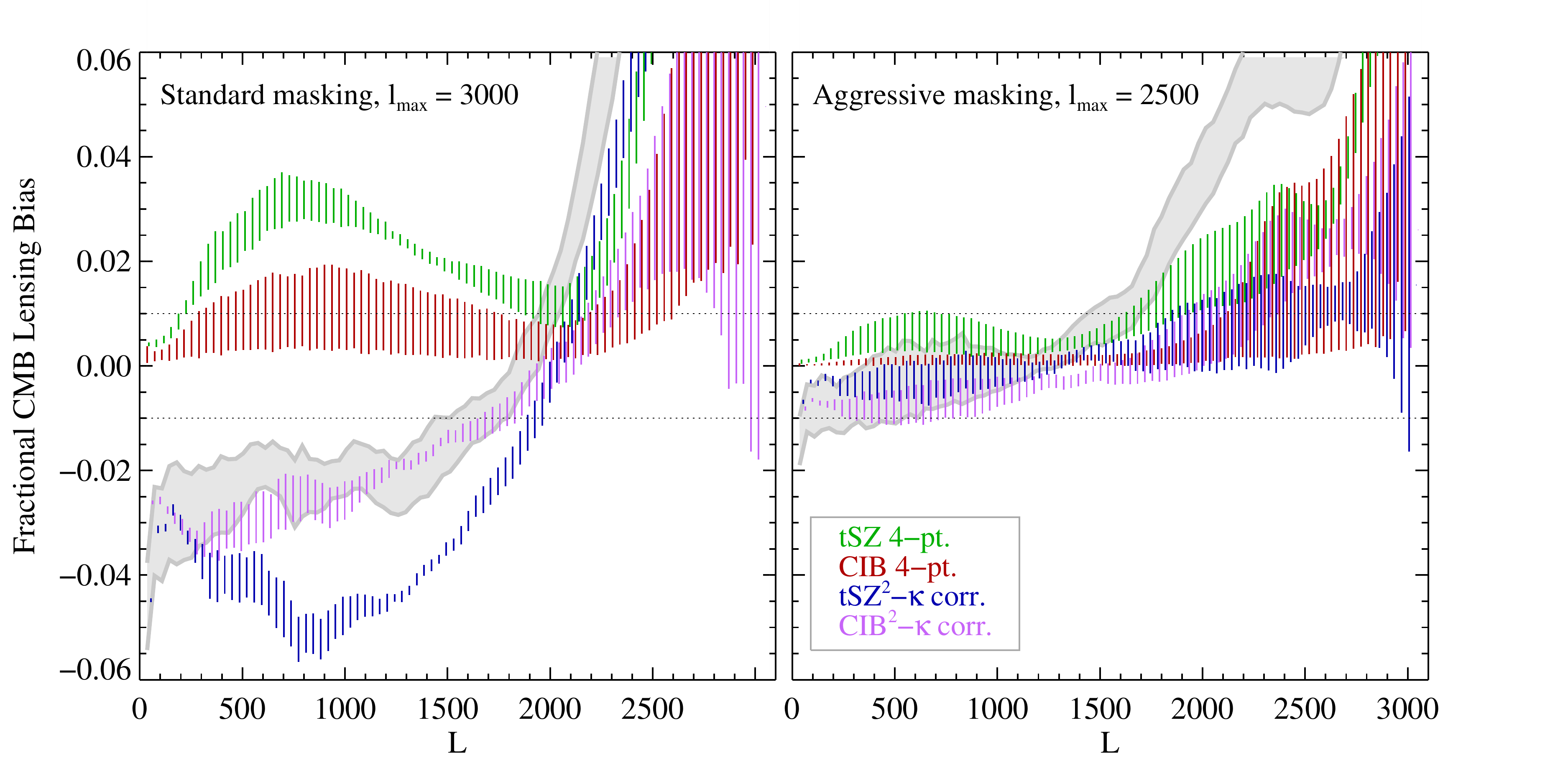}
  \caption{Summary of fractional CMB lensing bias levels from our simulations, as a function of lensing multipole, L.  Each colored region indicates the range of mean biases from the simulations we used.   Green spans the thermal SZ four-point biases, red spans the CIB four-point bias, blue spans the bias from the tSZ-$\kappa$ correlation, and pink spans the bias from the CIB-$\kappa$ correlation.  The range of total biases is bounded by the grey region, including the tSZ-CIB correlation found in the simulations, which slightly reduces the total bias.  A 1\% error band is indicated by the dotted  lines.  The left panel corresponds to masking sources above $5\,$mJy and clusters above $\mvir = 5\e{14}\,\Msol$.  The right panel corresponds to aggressive masking, with sources masked above 1\,mJy and clusters above $\mvir = 10^{14}\,\Msol$.  In the right panel, we also reduce the maximum temperature multipole used in the reconstruction to $\lmax = 2500$.  The region where the total is within 1\%,  $L<1400$,  accounts for more than $99.9\%$ of the total squared lensing signal-to-noise ratio.}
  \label{fig:total_components}
\end{figure*}

\section{Discussion}
In this paper we have studied several sources of bias on reconstructed CMB lensing power spectra that originate from known sources of non-Gaussianity in the millimeter-wave sky on small angular scales.  

Lensing analyses using high angular-resolution maps, such as those from ACT and SPT, yield much stronger lensing detections per unit sky area than analyses of maps from a lower-resolution experiment such as \Planck.  Intuitively this is due to the increased number of CMB mode pairs over which to average, out to the higher effective maximum multipole $\lmax$.  However, at high $\lmax$, foreground fluctuations also become increasingly important, to the point that they dominate the observed power spectrum  at $l \gtrsim 3000$.  To date, the analysis of temperature maps from ACT  (\citealt{das11b,das13}) and SPT \citep{vanengelen12} has yielded lensing detections at low enough significance that these biases could be neglected, with the smallest uncertainty on the lensing amplitude to date being the 16\% of  \citet{vanengelen12}.    However, current and upcoming analyses  will map sky areas which are larger by factors of several than these, and possibly with lower noise levels (in the case of a wide survey with ACTpol).  With statistical uncertainties of a few percent on the lensing amplitude, systematic effects need to be understood and controlled,  ideally to  a percent or better.


Point sources can be detected to the relatively low flux levels of several mJy in  maps such as those from ACT and SPT, particularly with the inclusion of data at multiple wavelengths.  If point sources are uncorrelated, a nonzero trispectrum impacts the inferred lensing amplitude, but this bias is sub-percent after applying standard masking thresholds.  In addition, the fact that the trispectrum is constant in multipole space for these sources means that this bias can be treated with other approaches, such as projecting it out of the reconstructed map \citep{namikawa13}.  

To treat other types of non-Gaussian foregrounds, particularly those with a different shape in multipole space, we analyzed  two independent, realistic sets of simulations (S10, B13).  For the CIB portions of these simulations, we first rescaled the amplitudes of the maps to match the observed power spectra.  We then estimated the bispectra for these simulations, finding   reasonable agreement with recent measures from SPT \citep{crawford13} and \Planck~\citep{planck13_p30}.  Performing lensing reconstructions on these fields, we isolated two types of bias; the first originates from the connected four-point function of the CIB, and the second originates from the correlation of the squared CIB with the lensing field.  Since these biases are of opposite sign there is some degree of cancellation.  We found that 
both sources of bias can impact the lensing amplitude at the level of several percent, with the latter type of bias being larger.  If masking is chosen as the method to treat this bias, we find that masking to $\sim 1\,$mJy achieves percent-level biases.




Fluctuations from the thermal Sunyaev-Zel'dovich effect can also lead to substantial biases, even when masking objects that are confidently detected.  We computed the biases from the  tSZ  simulations of S10 and B13, both of which contain updated  gas models designed to match the recent measurements of the power spectrum of tSZ fluctuations.  Again we found that for standard masking levels, biases of a few percent can remain, though there is some cancellation between the two types.

We explored the uncertainty in the tSZ trispectrum, originating from its dependence on the details of the cluster gas profiles and the cosmological model.   Using an analytical model of the tSZ trispectrum on the scales of relevance for CMB lensing, we calculated the four point-induced bias.  We then perturbed in the space of cosmological and cluster-physics parameters, the parameters which most affect the inferred lensing bias, leading to a large uncertainty.  It thus seems necessary to use either aggressive cluster masking, input from other frequencies, or an estimation of the tSZ trispectrum from the map itself to reduce this bias to percent levels.

Figure \ref{fig:total_components} summarizes our simulation-derived findings, with bands indicating the spread between the mean lensing biases for the various models.  The left panel corresponds to ``standard'' masking, with clusters and sources masked to their approximate $5\sigma$ thresholds of $\smax = 5\,$mJy and $\mvir = 5\e{14}\, \msol$.  The right panel corresponds to an ``aggressive'' masking level of $\smax = 1\,$mJy and $\mvir = 1\e{14} \msol$.  In order to reduce the total bias on the lensing power spectrum to be less than one percent, we have additionally found it necessary to reduce the maximal temperature multipole in the lens reconstruction from $\lmax = 3000$ to $\lmax = 2500$ for this case.  In this figure we have also allowed for the anticorrelation between the CIB and tSZ at 150~GHz \citep{addison12b}, which reduces the overall biases by $\sim 20\%$.

Given the strong dependence of the biases on source masking levels, we also studied the feasibility of masking very aggressively.  We used an in-painting routine to fill in the lensed CMB and noise fluctuations at the locations of masked sources and clusters, finding a negative bias on the reconstructed lensing power that roughly scales with the fraction of sky masked, for sufficiently small masked sky fractions.
Attempting to reduce this bias by in-painting in the reconstructed lensing  fields reduces this bias in a way that depends sensitively on the size of the in-painted region.  For an experimental analysis, it may be necessary to include this in-painting bias as a transferring effect on the reconstructed lensing power using detailed simulations, as we have done here.

The next few years promise to be an exciting time for CMB lensing, as it matures into a precision cosmological probe.  Here we have shown that although the biases from foregrounds are not trivial, they do not present an insurmountable obstacle to using CMB lensing as a new, clean probe for physics that affects the growth of structure.

\acknowledgments
We thank Blake Sherwin, Duncan Hanson,  Sudeep Das, and Olivier Dor\'{e} for useful discussions.  We thank Katrin Heitmann for providing the Coyote simulation run; Thibaut Louis for discussions as well as providing the in-painting code used in Section~\ref{sec:masking}; Tom Crawford for providing code used to process simulations; and Tijmen de Haan for computing support.  We are aware of closely-related work by Osbourne et al.~that is about to appear and thank those authors for useful discussion.
This work was supported by CIfAR, NSERC, and the Canada Research Chairs program.

\bibliography{refs.bib}

\begin{thebibliography}{89}
\expandafter\ifx\csname natexlab\endcsname\relax\def\natexlab#1{#1}\fi

\bibitem[{{Addison} {et~al.}(2012){Addison}, {Dunkley}, \&
  {Spergel}}]{addison12b}
{Addison}, G.~E., {Dunkley}, J., \& {Spergel}, D.~N. 2012, \mnras, 427, 1741

\bibitem[{{Amblard} {et~al.}(2004){Amblard}, {Vale}, \& {White}}]{amblard04}
{Amblard}, A., {Vale}, C., \& {White}, M. 2004, New Astonomy, 9, 687

\bibitem[{{Arg{\"u}eso} {et~al.}(2003){Arg{\"u}eso}, {Gonz{\'a}lez-Nuevo}, \&
  {Toffolatti}}]{argueso03}
{Arg{\"u}eso}, F., {Gonz{\'a}lez-Nuevo}, J., \& {Toffolatti}, L. 2003, \apj,
  598, 86

\bibitem[{{Bai} {et~al.}(2007){Bai}, {Marcillac}, {Rieke}, {Rieke}, {Tran},
  {Hinz}, {Rudnick}, {Kelly}, \& {Blaylock}}]{bai07}
{Bai}, L., {et~al.} 2007, \apj, 664, 181

\bibitem[{{Battaglia} {et~al.}(2010){Battaglia}, {Bond}, {Pfrommer}, {Sievers},
  \& {Sijacki}}]{battaglia10}
{Battaglia}, N., {Bond}, J.~R., {Pfrommer}, C., {Sievers}, J.~L., \& {Sijacki},
  D. 2010, \apj, 725, 91

\bibitem[{{Benoit-L{\'e}vy} {et~al.}(2013){Benoit-L{\'e}vy}, {D{\'e}chelette},
  {Benabed}, {Cardoso}, {Hanson}, \& {Prunet}}]{benoitlevy13}
{Benoit-L{\'e}vy}, A., {D{\'e}chelette}, T., {Benabed}, K., {Cardoso}, J.-F.,
  {Hanson}, D., \& {Prunet}, S. 2013, ArXiv e-prints, 1301.4145

\bibitem[{{Benson} {et~al.}(2013){Benson}, {de Haan}, {Dudley}, {Reichardt},
  {Aird}, {Andersson}, {Armstrong}, {Ashby}, {Bautz}, {Bayliss}, {Bazin},
  {Bleem}, {Brodwin}, {Carlstrom}, {Chang}, {Cho}, {Clocchiatti}, {Crawford},
  {Crites}, {Desai}, {Dobbs}, {Foley}, {Forman}, {George}, {Gladders},
  {Gonzalez}, {Halverson}, {Harrington}, {High}, {Holder}, {Holzapfel},
  {Hoover}, {Hrubes}, {Jones}, {Joy}, {Keisler}, {Knox}, {Lee}, {Leitch},
  {Liu}, {Lueker}, {Luong-Van}, {Mantz}, {Marrone}, {McDonald}, {McMahon},
  {Mehl}, {Meyer}, {Mocanu}, {Mohr}, {Montroy}, {Murray}, {Natoli}, {Padin},
  {Plagge}, {Pryke}, {Rest}, {Ruel}, {Ruhl}, {Saliwanchik}, {Saro}, {Sayre},
  {Schaffer}, {Shaw}, {Shirokoff}, {Song}, {Spieler}, {Stalder},
  {Staniszewski}, {Stark}, {Story}, {Stubbs}, {Suhada}, {van Engelen},
  {Vanderlinde}, {Vieira}, {Vikhlinin}, {Williamson}, {Zahn}, \&
  {Zenteno}}]{benson13}
{Benson}, B.~A., {et~al.} 2013, \apj, 763, 147

\bibitem[{{Bernardeau}(1997)}]{bernardeau97}
{Bernardeau}, F. 1997, \aap, 324, 15

\bibitem[{{B{\'e}thermin} {et~al.}(2011){B{\'e}thermin}, {Dole}, {Lagache}, {Le
  Borgne}, \& {Penin}}]{bethermin11}
{B{\'e}thermin}, M., {Dole}, H., {Lagache}, G., {Le Borgne}, D., \& {Penin}, A.
  2011, \aap, 529, A4+

\bibitem[{{Bhattacharya} {et~al.}(2012){Bhattacharya}, {Nagai}, {Shaw},
  {Crawford}, \& {Holder}}]{bhattacharya12}
{Bhattacharya}, S., {Nagai}, D., {Shaw}, L., {Crawford}, T., \& {Holder}, G.~P.
  2012, ArXiv e-prints, 1203.6368

\bibitem[{{Bleem} {et~al.}(2012){Bleem}, {van Engelen}, {Holder}, {Aird},
  {Armstrong}, {Ashby}, {Becker}, {Benson}, {Biesiadzinski}, {Brodwin},
  {Busha}, {Carlstrom}, {Chang}, {Cho}, {Crawford}, {Crites}, {de Haan},
  {Desai}, {Dobbs}, {Dor{\'e}}, {Dudley}, {Geach}, {George}, {Gladders},
  {Gonzalez}, {Halverson}, {Harrington}, {High}, {Holden}, {Holzapfel},
  {Hoover}, {Hrubes}, {Joy}, {Keisler}, {Knox}, {Lee}, {Leitch}, {Lueker},
  {Luong-Van}, {Marrone}, {Martinez-Manso}, {McMahon}, {Mehl}, {Meyer}, {Mohr},
  {Montroy}, {Natoli}, {Padin}, {Plagge}, {Pryke}, {Reichardt}, {Rest}, {Ruhl},
  {Saliwanchik}, {Sayre}, {Schaffer}, {Shaw}, {Shirokoff}, {Spieler},
  {Stalder}, {Stanford}, {Staniszewski}, {Stark}, {Stern}, {Story},
  {Vallinotto}, {Vanderlinde}, {Vieira}, {Wechsler}, {Williamson}, \&
  {Zahn}}]{bleem12b}
{Bleem}, L.~E., {et~al.} 2012, \apj, 753, L9

\bibitem[{{Bode} {et~al.}(2012){Bode}, {Ostriker}, {Cen}, \& {Trac}}]{bode12}
{Bode}, P., {Ostriker}, J.~P., {Cen}, R., \& {Trac}, H. 2012, ArXiv e-prints,
  1204.1762

\bibitem[{{Bucher} \& {Louis}(2012)}]{bucher12}
{Bucher}, M., \& {Louis}, T. 2012, \mnras, 424, 1694

\bibitem[{{Carlstrom} {et~al.}(2002){Carlstrom}, {Holder}, \&
  {Reese}}]{carlstrom02}
{Carlstrom}, J.~E., {Holder}, G.~P., \& {Reese}, E.~D. 2002, \araa, 40, 643

\bibitem[{{Challinor} \& {Lewis}(2005)}]{challinor05}
{Challinor}, A., \& {Lewis}, A. 2005, \prd, 71, 103010

\bibitem[{{Cooray}(2001)}]{cooray01b}
{Cooray}, A. 2001, \prd, 64, 063514

\bibitem[{{Cooray} \& {Kesden}(2003)}]{cooray03}
{Cooray}, A., \& {Kesden}, M. 2003, New Astronomy, 8, 231

\bibitem[{{Coppin} {et~al.}(2005){Coppin}, {Halpern}, {Scott}, {Borys}, \&
  {Chapman}}]{coppin05}
{Coppin}, K., {Halpern}, M., {Scott}, D., {Borys}, C., \& {Chapman}, S. 2005,
  \mnras, 357, 1022

\bibitem[{{Crawford} {et~al.}(2013){Crawford}, {Schaffer}, {Bhattacharya},
  {Aird}, {Benson}, {Bleem}, {Carlstrom}, {Chang}, {Cho}, {Crites}, {de Haan},
  {Dobbs}, {Dudley}, {George}, {Halverson}, {Holder}, {Holzapfel}, {Hoover},
  {Hou}, {Hrubes}, {Keisler}, {Knox}, {Lee}, {Leitch}, {Lueker}, {Luong-Van},
  {McMahon}, {Mehl}, {Meyer}, {Millea}, {Mocanu}, {Mohr}, {Montroy}, {Padin},
  {Plagge}, {Pryke}, {Reichardt}, {Ruhl}, {Sayre}, {Shaw}, {Shirokoff},
  {Spieler}, {Staniszewski}, {Stark}, {Story}, {van Engelen}, {Vanderlinde},
  {Vieira}, {Williamson}, \& {Zahn}}]{crawford13}
{Crawford}, T.~M., {et~al.} 2013, ArXiv e-prints, 1303.3535

\bibitem[{{Das} {et~al.}(2013){Das}, {Louis}, {Nolta}, {Addison},
  {Battistelli}, {Bond}, {Calabrese}, {Devlin}, {Dicker}, {Dunkley},
  {D{\"u}nner}, {Fowler}, {Gralla}, {Hajian}, {Halpern}, {Hasselfield},
  {Hilton}, {Hincks}, {Hlozek}, {Huffenberger}, {Hughes}, {Irwin}, {Kosowsky},
  {Lupton}, {Marriage}, {Marsden}, {Menanteau}, {Moodley}, {Niemack}, {Page},
  {Partridge}, {Reese}, {Schmitt}, {Sehgal}, {Sherwin}, {Sievers}, {Spergel},
  {Staggs}, {Swetz}, {Switzer}, {Thornton}, {Trac}, \& {Wollack}}]{das13}
{Das}, S., {et~al.} 2013, ArXiv e-prints, 1301.1037

\bibitem[{{Das} {et~al.}(2011){Das}, {Marriage}, {Ade}, {Aguirre}, {Amiri},
  {Appel}, {Barrientos}, {Battistelli}, {Bond}, {Brown}, {Burger}, {Chervenak},
  {Devlin}, {Dicker}, {Bertrand Doriese}, {Dunkley}, {D{\"u}nner},
  {Essinger-Hileman}, {Fisher}, {Fowler}, {Hajian}, {Halpern}, {Hasselfield},
  {Hern{\'a}ndez-Monteagudo}, {Hilton}, {Hilton}, {Hincks}, {Hlozek},
  {Huffenberger}, {Hughes}, {Hughes}, {Infante}, {Irwin}, {Baptiste Juin},
  {Kaul}, {Klein}, {Kosowsky}, {Lau}, {Limon}, {Lin}, {Lupton}, {Marsden},
  {Martocci}, {Mauskopf}, {Menanteau}, {Moodley}, {Moseley}, {Netterfield},
  {Niemack}, {Nolta}, {Page}, {Parker}, {Partridge}, {Reid}, {Sehgal},
  {Sherwin}, {Sievers}, {Spergel}, {Staggs}, {Swetz}, {Switzer}, {Thornton},
  {Trac}, {Tucker}, {Warne}, {Wollack}, \& {Zhao}}]{das11b}
------. 2011, \apj, 729, 62

\bibitem[{{de Zotti} {et~al.}(2005){de Zotti}, {Ricci}, {Mesa}, {Silva},
  {Mazzotta}, {Toffolatti}, \& {Gonz{\'a}lez-Nuevo}}]{dezotti05}
{de Zotti}, G., {Ricci}, R., {Mesa}, D., {Silva}, L., {Mazzotta}, P.,
  {Toffolatti}, L., \& {Gonz{\'a}lez-Nuevo}, J. 2005, \aap, 431, 893

\bibitem[{{Dunkley} {et~al.}(2011){Dunkley}, {Hlozek}, {Sievers}, {Acquaviva},
  {Ade}, {Aguirre}, {Amiri}, {Appel}, {Barrientos}, {Battistelli}, {Bond},
  {Brown}, {Burger}, {Chervenak}, {Das}, {Devlin}, {Dicker}, {Bertrand
  Doriese}, {D{\"u}nner}, {Essinger-Hileman}, {Fisher}, {Fowler}, {Hajian},
  {Halpern}, {Hasselfield}, {Hern{\'a}ndez-Monteagudo}, {Hilton}, {Hilton},
  {Hincks}, {Huffenberger}, {Hughes}, {Hughes}, {Infante}, {Irwin}, {Juin},
  {Kaul}, {Klein}, {Kosowsky}, {Lau}, {Limon}, {Lin}, {Lupton}, {Marriage},
  {Marsden}, {Mauskopf}, {Menanteau}, {Moodley}, {Moseley}, {Netterfield},
  {Niemack}, {Nolta}, {Page}, {Parker}, {Partridge}, {Reid}, {Sehgal},
  {Sherwin}, {Spergel}, {Staggs}, {Swetz}, {Switzer}, {Thornton}, {Trac},
  {Tucker}, {Warne}, {Wollack}, \& {Zhao}}]{dunkley11}
{Dunkley}, J., {et~al.} 2011, \apj, 739, 52

\bibitem[{{Dvorkin} \& {Smith}(2009)}]{dvorkinsmith09}
{Dvorkin}, C., \& {Smith}, K.~M. 2009, \prd, 79, 043003

\bibitem[{{Fixsen} {et~al.}(1998){Fixsen}, {Dwek}, {Mather}, {Bennett}, \&
  {Shafer}}]{fixsen98}
{Fixsen}, D.~J., {Dwek}, E., {Mather}, J.~C., {Bennett}, C.~L., \& {Shafer},
  R.~A. 1998, \apj, 508, 123

\bibitem[{{Hall} {et~al.}(2010){Hall}, {Knox}, {Reichardt}, {Ade}, {Aird},
  {Benson}, {Bleem}, {Carlstrom}, {Chang}, {Cho}, {Crawford}, {Crites}, {de
  Haan}, {Dobbs}, {George}, {Halverson}, {Holder}, {Holzapfel}, {Hrubes},
  {Joy}, {Keisler}, {Lee}, {Leitch}, {Lueker}, {McMahon}, {Mehl}, {Meyer},
  {Mohr}, {Montroy}, {Padin}, {Plagge}, {Pryke}, {Ruhl}, {Schaffer}, {Shaw},
  {Shirokoff}, {Spieler}, {Staniszewski}, {Stark}, {Switzer}, {Vanderlinde},
  {Vieira}, {Williamson}, \& {Zahn}}]{hall10}
{Hall}, N.~R., {et~al.} 2010, \apj, 718, 632

\bibitem[{{Hanson} {et~al.}(2011){Hanson}, {Challinor}, {Efstathiou}, \&
  {Bielewicz}}]{hanson11}
{Hanson}, D., {Challinor}, A., {Efstathiou}, G., \& {Bielewicz}, P. 2011, \prd,
  83, 043005

\bibitem[{{Hanson} {et~al.}(2013){Hanson}, {Hoover}, {Crites}, {Ade}, {Aird},
  {Austermann}, {Beall}, {Bender}, {Benson}, {Bleem}, {Bock}, {Carlstrom},
  {Chang}, {Chiang}, {Cho}, {Conley}, {Crawford}, {de Haan}, {Dobbs},
  {Everett}, {Gallicchio}, {Gao}, {George}, {Halverson}, {Harrington},
  {Henning}, {Hilton}, {Holder}, {Holzapfel}, {Hrubes}, {Huang}, {Hubmayr},
  {Irwin}, {Keisler}, {Knox}, {Lee}, {Leitch}, {Li}, {Liang}, {Luong-Van},
  {Marsden}, {McMahon}, {Mehl}, {Meyer}, {Mocanu}, {Montroy}, {Natoli},
  {Nibarger}, {Novosad}, {Padin}, {Pryke}, {Reichardt}, {Ruhl}, {Saliwanchik},
  {Sayre}, {Schaffer}, {Schulz}, {Smecher}, {Stark}, {Story}, {Tucker},
  {Vanderlinde}, {Vieira}, {Viero}, {Wang}, {Yefremenko}, {Zahn}, \&
  {Zemcov}}]{hanson13}
{Hanson}, D., {et~al.} 2013, Physical Review Letters, 111, 141301

\bibitem[{{Hanson} {et~al.}(2009){Hanson}, {Rocha}, \& {G{\'o}rski}}]{hanson09}
{Hanson}, D., {Rocha}, G., \& {G{\'o}rski}, K. 2009, \mnras, 400, 2169

\bibitem[{{Hasselfield} {et~al.}(2013){Hasselfield}, {Hilton}, {Marriage},
  {Addison}, {Barrientos}, {Battaglia}, {Battistelli}, {Bond}, {Crichton},
  {Das}, {Devlin}, {Dicker}, {Dunkley}, {D{\"u}nner}, {Fowler}, {Gralla},
  {Hajian}, {Halpern}, {Hincks}, {Hlozek}, {Hughes}, {Infante}, {Irwin},
  {Kosowsky}, {Marsden}, {Menanteau}, {Moodley}, {Niemack}, {Nolta}, {Page},
  {Partridge}, {Reese}, {Schmitt}, {Sehgal}, {Sherwin}, {Sievers}, {Sif{\'o}n},
  {Spergel}, {Staggs}, {Swetz}, {Switzer}, {Thornton}, {Trac}, \&
  {Wollack}}]{hasslefield13}
{Hasselfield}, M., {et~al.} 2013, Journal of Cosmology and Astroparticle
  Physics, 7, 8

\bibitem[{{Heitmann} {et~al.}(2010){Heitmann}, {White}, {Wagner}, {Habib}, \&
  {Higdon}}]{heitmann10}
{Heitmann}, K., {White}, M., {Wagner}, C., {Habib}, S., \& {Higdon}, D. 2010,
  \apj, 715, 104

\bibitem[{{Hill} \& {Sherwin}(2013)}]{hill13}
{Hill}, J.~C., \& {Sherwin}, B.~D. 2013, \prd, 87, 023527

\bibitem[{{Hoffman} \& {Ribak}(1991)}]{hoffman91}
{Hoffman}, Y., \& {Ribak}, E. 1991, \apjl, 380, L5

\bibitem[{{Holder} {et~al.}(2013){Holder}, {Viero}, {Zahn}, {Aird}, {Benson},
  {Bhattacharya}, {Bleem}, {Bock}, {Brodwin}, {Carlstrom}, {Chang}, {Cho},
  {Conley}, {Crawford}, {Crites}, {de Haan}, {Dobbs}, {Dudley}, {George},
  {Halverson}, {Holzapfel}, {Hoover}, {Hou}, {Hrubes}, {Keisler}, {Knox},
  {Lee}, {Leitch}, {Lueker}, {Luong-Van}, {Marsden}, {Marrone}, {McMahon},
  {Mehl}, {Meyer}, {Millea}, {Mohr}, {Montroy}, {Padin}, {Plagge}, {Pryke},
  {Reichardt}, {Ruhl}, {Sayre}, {Schaffer}, {Schulz}, {Shaw}, {Shirokoff},
  {Spieler}, {Staniszewski}, {Stark}, {Story}, {van Engelen}, {Vanderlinde},
  {Vieira}, {Williamson}, \& {Zemcov}}]{holder13}
{Holder}, G.~P., {et~al.} 2013, \apjl, 771, L16

\bibitem[{{Hu}(2000)}]{hu00}
{Hu}, W. 2000, \apj, 529, 12

\bibitem[{{Hu}(2001)}]{hu01b}
------. 2001, \apjl, 557, L79

\bibitem[{{Hu} \& {Okamoto}(2002)}]{hu02a}
{Hu}, W., \& {Okamoto}, T. 2002, \apj, 574, 566

\bibitem[{{Keisler} {et~al.}(2011){Keisler}, {Reichardt}, {Aird}, {Benson},
  {Bleem}, {Carlstrom}, {Chang}, {Cho}, {Crawford}, {Crites}, {de Haan},
  {Dobbs}, {Dudley}, {George}, {Halverson}, {Holder}, {Holzapfel}, {Hoover},
  {Hou}, {Hrubes}, {Joy}, {Knox}, {Lee}, {Leitch}, {Lueker}, {Luong-Van},
  {McMahon}, {Mehl}, {Meyer}, {Millea}, {Mohr}, {Montroy}, {Natoli}, {Padin},
  {Plagge}, {Pryke}, {Ruhl}, {Schaffer}, {Shaw}, {Shirokoff}, {Spieler},
  {Staniszewski}, {Stark}, {Story}, {van Engelen}, {Vanderlinde}, {Vieira},
  {Williamson}, \& {Zahn}}]{keisler11}
{Keisler}, R., {et~al.} 2011, \apj, 743, 28

\bibitem[{{Kesden} {et~al.}(2003){Kesden}, {Cooray}, \&
  {Kamionkowski}}]{kesden03}
{Kesden}, M., {Cooray}, A., \& {Kamionkowski}, M. 2003, \prd, 67, 123507

\bibitem[{{Komatsu} \& {Kitayama}(1999)}]{komatsu99a}
{Komatsu}, E., \& {Kitayama}, T. 1999, \apjl, 526, L1

\bibitem[{{Komatsu} \& {Seljak}(2002)}]{komatsu02}
{Komatsu}, E., \& {Seljak}, U. 2002, \mnras, 336, 1256

\bibitem[{{Komatsu} {et~al.}(2011){Komatsu}, {Smith}, {Dunkley}, {Bennett},
  {Gold}, {Hinshaw}, {Jarosik}, {Larson}, {Nolta}, {Page}, {Spergel},
  {Halpern}, {Hill}, {Kogut}, {Limon}, {Meyer}, {Odegard}, {Tucker}, {Weiland},
  {Wollack}, \& {Wright}}]{Komatsu11}
{Komatsu}, E., {et~al.} 2011, \apjs, 192, 18

\bibitem[{{Lacasa} {et~al.}(2012){Lacasa}, {Aghanim}, {Kunz}, \&
  {Frommert}}]{lacasa12}
{Lacasa}, F., {Aghanim}, N., {Kunz}, M., \& {Frommert}, M. 2012, \mnras, 421,
  1982

\bibitem[{{Lesgourgues} {et~al.}(2005){Lesgourgues}, {Liguori}, {Matarrese}, \&
  {Riotto}}]{lesgourgues05}
{Lesgourgues}, J., {Liguori}, M., {Matarrese}, S., \& {Riotto}, A. 2005, \prd,
  71, 103514

\bibitem[{{Lewis} \& {Challinor}(2006)}]{lewis06}
{Lewis}, A., \& {Challinor}, A. 2006, \physrep, 429, 1

\bibitem[{{Lueker} {et~al.}(2010){Lueker}, {Reichardt}, {Schaffer}, {Zahn},
  {Ade}, {Aird}, {Benson}, {Bleem}, {Carlstrom}, {Chang}, {Cho}, {Crawford},
  {Crites}, {de Haan}, {Dobbs}, {George}, {Hall}, {Halverson}, {Holder},
  {Holzapfel}, {Hrubes}, {Joy}, {Keisler}, {Knox}, {Lee}, {Leitch}, {McMahon},
  {Mehl}, {Meyer}, {Mohr}, {Montroy}, {Padin}, {Plagge}, {Pryke}, {Ruhl},
  {Shaw}, {Shirokoff}, {Spieler}, {Stalder}, {Staniszewski}, {Stark},
  {Vanderlinde}, {Vieira}, \& {Williamson}}]{lueker10}
{Lueker}, M., {et~al.} 2010, \apj, 719, 1045

\bibitem[{{Marsden} {et~al.}(2013){Marsden}, {Gralla}, {Marriage}, {Switzer},
  {Partridge}, {Massardi}, {Morales}, {Addison}, {Bond}, {Crichton}, {Das},
  {Devlin}, {Dunner}, {Hajian}, {Hilton}, {Hincks}, {Hughes}, {Irwin},
  {Kosowsky}, {Menanteau}, {Moodley}, {Niemack}, {Page}, {Reese}, {Schmitt},
  {Sehgal}, {Sievers}, {Staggs}, {Swetz}, {Thornton}, \& {Wollack}}]{marsden13}
{Marsden}, D., {et~al.} 2013, ArXiv e-prints, 1306.2288

\bibitem[{{McMahon} {et~al.}(2009){McMahon}, {Aird}, {Benson}, {Bleem},
  {Britton}, {Carlstrom}, {Chang}, {Cho}, {de Haan}, {Crawford}, {Crites},
  {Datesman}, {Dobbs}, {Everett}, {Halverson}, {Holder}, {Holzapfel}, {Hrubes},
  {Irwin}, {Joy}, {Keisler}, {Lanting}, {Lee}, {Leitch}, {Loehr}, {Lueker},
  {Mehl}, {Meyer}, {Mohr}, {Montroy}, {Niemack}, {Ngeow}, {Novosad}, {Padin},
  {Plagge}, {Pryke}, {Reichardt}, {Ruhl}, {Schaffer}, {Shaw}, {Shirokoff},
  {Spieler}, {Stadler}, {Stark}, {Staniszewski}, {Vanderlinde}, {Vieira},
  {Wang}, {Williamson}, {Yefremenko}, {Yoon}, {Zhan}, \& {Zenteno}}]{mcmahon09}
{McMahon}, J.~J., {et~al.} 2009, in American Institute of Physics Conference
  Series, Vol. 1185, American Institute of Physics Conference Series, ed.
  {B.~Young, B.~Cabrera, \& A.~Miller}, 511--514

\bibitem[{{Merkel} \& {Sch{\"a}fer}(2013)}]{merkel13}
{Merkel}, P.~M., \& {Sch{\"a}fer}, B.~M. 2013, \mnras, 429, 444

\bibitem[{{Mocanu} {et~al.}(2013){Mocanu}, {Crawford}, {Vieira}, {Aird},
  {Aravena}, {Austermann}, {Benson}, {B{\'e}thermin}, {Bleem}, {Bothwell},
  {Carlstrom}, {Chang}, {Chapman}, {Cho}, {Crites}, {de Haan}, {Dobbs},
  {Everett}, {George}, {Halverson}, {Harrington}, {Hezaveh}, {Holder},
  {Holzapfel}, {Hoover}, {Hrubes}, {Keisler}, {Knox}, {Lee}, {Leitch},
  {Lueker}, {Luong-Van}, {Marrone}, {McMahon}, {Mehl}, {Meyer}, {Mohr},
  {Montroy}, {Natoli}, {Padin}, {Plagge}, {Pryke}, {Rest}, {Reichardt}, {Ruhl},
  {Sayre}, {Schaffer}, {Shirokoff}, {Spieler}, {Spilker}, {Stalder},
  {Staniszewski}, {Stark}, {Story}, {Switzer}, {Vanderlinde}, \&
  {Williamson}}]{mocanu13}
{Mocanu}, L.~M., {et~al.} 2013, ArXiv e-prints, 1306.3470

\bibitem[{{Namikawa} {et~al.}(2012){Namikawa}, {Hanson}, \&
  {Takahashi}}]{namikawa12}
{Namikawa}, T., {Hanson}, D., \& {Takahashi}, R. 2012, ArXiv e-prints,
  1209.0091

\bibitem[{{Namikawa} \& {Takahashi}(2013)}]{namikawa13}
{Namikawa}, T., \& {Takahashi}, R. 2013, ArXiv e-prints, 1310.2372

\bibitem[{{Navarro} {et~al.}(1997){Navarro}, {Frenk}, \& {White}}]{navarro97}
{Navarro}, J.~F., {Frenk}, C.~S., \& {White}, S.~D.~M. 1997, \apj, 490, 493

\bibitem[{{Negrello} {et~al.}(2007){Negrello}, {Perrotta},
  {Gonz{\'a}lez-Nuevo}, {Silva}, {de Zotti}, {Granato}, {Baccigalupi}, \&
  {Danese}}]{negrello07}
{Negrello}, M., {Perrotta}, F., {Gonz{\'a}lez-Nuevo}, J., {Silva}, L., {de
  Zotti}, G., {Granato}, G.~L., {Baccigalupi}, C., \& {Danese}, L. 2007,
  \mnras, 377, 1557

\bibitem[{{Niemack} {et~al.}(2010){Niemack}, {Ade}, {Aguirre}, {Barrientos},
  {Beall}, {Bond}, {Britton}, {Cho}, {Das}, {Devlin}, {Dicker}, {Dunkley},
  {D{\"u}nner}, {Fowler}, {Hajian}, {Halpern}, {Hasselfield}, {Hilton},
  {Hilton}, {Hubmayr}, {Hughes}, {Infante}, {Irwin}, {Jarosik}, {Klein},
  {Kosowsky}, {Marriage}, {McMahon}, {Menanteau}, {Moodley}, {Nibarger},
  {Nolta}, {Page}, {Partridge}, {Reese}, {Sievers}, {Spergel}, {Staggs},
  {Thornton}, {Tucker}, {Wollack}, \& {Yoon}}]{niemack10}
{Niemack}, M.~D., {et~al.} 2010, in Society of Photo-Optical Instrumentation
  Engineers (SPIE) Conference Series, Vol. 7741, Society of Photo-Optical
  Instrumentation Engineers (SPIE) Conference Series

\bibitem[{{Perotto} {et~al.}(2010){Perotto}, {Bobin}, {Plaszczynski}, {Starck},
  \& {Lavabre}}]{perotto10}
{Perotto}, L., {Bobin}, J., {Plaszczynski}, S., {Starck}, J., \& {Lavabre}, A.
  2010, \aap, 519, A4+

\bibitem[{{Planck Collaboration} {et~al.}(2013{\natexlab{a}}){Planck
  Collaboration}, {Ade}, {Aghanim}, {Armitage-Caplan}, {Arnaud}, {Ashdown},
  {Atrio-Barandela}, {Aumont}, {Baccigalupi}, {Banday}, \&
  et~al.}]{planck13_p16}
{Planck Collaboration}, {et~al.} 2013{\natexlab{a}}, ArXiv e-prints, 1303.5076

\bibitem[{{Planck Collaboration} {et~al.}(2013{\natexlab{b}}){Planck
  Collaboration}, {Ade}, {Aghanim}, {Armitage-Caplan}, {Arnaud}, {Ashdown},
  {Atrio-Barandela}, {Aumont}, {Baccigalupi}, {Banday}, \&
  et~al.}]{planck13_p17}
------. 2013{\natexlab{b}}, ArXiv e-prints, 1303.5077

\bibitem[{{Planck Collaboration} {et~al.}(2013{\natexlab{c}}){Planck
  Collaboration}, {Ade}, {Aghanim}, {Armitage-Caplan}, {Arnaud}, {Ashdown},
  {Atrio-Barandela}, {Aumont}, {Baccigalupi}, {Banday}, \&
  et~al.}]{planck13_p18}
------. 2013{\natexlab{c}}, ArXiv e-prints, 1303.5078

\bibitem[{{Planck Collaboration} {et~al.}(2013{\natexlab{d}}){Planck
  Collaboration}, {Ade}, {Aghanim}, {Armitage-Caplan}, {Arnaud}, {Ashdown},
  {Atrio-Barandela}, {Aumont}, {Baccigalupi}, {Banday}, \&
  et~al.}]{planck13_p21}
------. 2013{\natexlab{d}}, ArXiv e-prints, 1303.5081

\bibitem[{{Planck Collaboration} {et~al.}(2013{\natexlab{e}}){Planck
  Collaboration}, {Ade}, {Aghanim}, {Armitage-Caplan}, {Arnaud}, {Ashdown},
  {Atrio-Barandela}, {Aumont}, {Baccigalupi}, {Banday}, \&
  et~al.}]{planck13_p30}
------. 2013{\natexlab{e}}, ArXiv e-prints, 1309.0382

\bibitem[{{Planck Collaboration} {et~al.}(2011){Planck Collaboration}, {Ade},
  {Aghanim}, {Arnaud}, {Ashdown}, {Aumont}, {Baccigalupi}, {Balbi}, {Banday},
  {Barreiro}, \& et~al.}]{planck11_p18}
------. 2011, \aap, 536, A18

\bibitem[{{Reichardt} {et~al.}(2009){Reichardt}, {Ade}, {Bock}, {Bond},
  {Brevik}, {Contaldi}, {Daub}, {Dempsey}, {Goldstein}, {Holzapfel}, {Kuo},
  {Lange}, {Lueker}, {Newcomb}, {Peterson}, {Ruhl}, {Runyan}, \&
  {Staniszewski}}]{reichardt09a}
{Reichardt}, C.~L., {et~al.} 2009, \apj, 694, 1200

\bibitem[{{Reichardt} {et~al.}(2012){Reichardt}, {Shaw}, {Zahn}, {Aird},
  {Benson}, {Bleem}, {Carlstrom}, {Chang}, {Cho}, {Crawford}, {Crites}, {de
  Haan}, {Dobbs}, {Dudley}, {George}, {Halverson}, {Holder}, {Holzapfel},
  {Hoover}, {Hou}, {Hrubes}, {Joy}, {Keisler}, {Knox}, {Lee}, {Leitch},
  {Lueker}, {Luong-Van}, {McMahon}, {Mehl}, {Meyer}, {Millea}, {Mohr},
  {Montroy}, {Natoli}, {Padin}, {Plagge}, {Pryke}, {Ruhl}, {Schaffer},
  {Shirokoff}, {Spieler}, {Staniszewski}, {Stark}, {Story}, {van Engelen},
  {Vanderlinde}, {Vieira}, \& {Williamson}}]{reichardt11}
------. 2012, \apj, 755, 70

\bibitem[{{Reichardt} {et~al.}(2013){Reichardt}, {Stalder}, {Bleem}, {Montroy},
  {Aird}, {Andersson}, {Armstrong}, {Ashby}, {Bautz}, {Bayliss}, {Bazin},
  {Benson}, {Brodwin}, {Carlstrom}, {Chang}, {Cho}, {Clocchiatti}, {Crawford},
  {Crites}, {de Haan}, {Desai}, {Dobbs}, {Dudley}, {Foley}, {Forman}, {George},
  {Gladders}, {Gonzalez}, {Halverson}, {Harrington}, {High}, {Holder},
  {Holzapfel}, {Hoover}, {Hrubes}, {Jones}, {Joy}, {Keisler}, {Knox}, {Lee},
  {Leitch}, {Liu}, {Lueker}, {Luong-Van}, {Mantz}, {Marrone}, {McDonald},
  {McMahon}, {Mehl}, {Meyer}, {Mocanu}, {Mohr}, {Murray}, {Natoli}, {Padin},
  {Plagge}, {Pryke}, {Rest}, {Ruel}, {Ruhl}, {Saliwanchik}, {Saro}, {Sayre},
  {Schaffer}, {Shaw}, {Shirokoff}, {Song}, {Spieler}, {Staniszewski}, {Stark},
  {Story}, {Stubbs}, {{\v S}uhada}, {van Engelen}, {Vanderlinde}, {Vieira},
  {Vikhlinin}, {Williamson}, {Zahn}, \& {Zenteno}}]{reichardt12}
------. 2013, \apj, 763, 127

\bibitem[{Rephaeli(1995)}]{rephaeli95}
Rephaeli, Y. 1995, \araa, 33, 541

\bibitem[{{Sehgal} {et~al.}(2010){Sehgal}, {Bode}, {Das},
  {Hernandez-Monteagudo}, {Huffenberger}, {Lin}, {Ostriker}, \&
  {Trac}}]{sehgal10}
{Sehgal}, N., {Bode}, P., {Das}, S., {Hernandez-Monteagudo}, C.,
  {Huffenberger}, K., {Lin}, Y., {Ostriker}, J.~P., \& {Trac}, H. 2010, \apj,
  709, 920

\bibitem[{{Sehgal} {et~al.}(2011){Sehgal}, {Trac}, {Acquaviva}, {Ade},
  {Aguirre}, {Amiri}, {Appel}, {Barrientos}, {Battistelli}, {Bond}, {Brown},
  {Burger}, {Chervenak}, {Das}, {Devlin}, {Dicker}, {Bertrand Doriese},
  {Dunkley}, {D{\"u}nner}, {Essinger-Hileman}, {Fisher}, {Fowler}, {Hajian},
  {Halpern}, {Hasselfield}, {Hern{\'a}ndez-Monteagudo}, {Hilton}, {Hilton},
  {Hincks}, {Hlozek}, {Holtz}, {Huffenberger}, {Hughes}, {Hughes}, {Infante},
  {Irwin}, {Jones}, {Baptiste Juin}, {Klein}, {Kosowsky}, {Lau}, {Limon},
  {Lin}, {Lupton}, {Marriage}, {Marsden}, {Martocci}, {Mauskopf}, {Menanteau},
  {Moodley}, {Moseley}, {Netterfield}, {Niemack}, {Nolta}, {Page}, {Parker},
  {Partridge}, {Reid}, {Sherwin}, {Sievers}, {Spergel}, {Staggs}, {Swetz},
  {Switzer}, {Thornton}, {Tucker}, {Warne}, {Wollack}, \& {Zhao}}]{sehgal11}
{Sehgal}, N., {et~al.} 2011, \apj, 732, 44

\bibitem[{{Seljak} \& {Zaldarriaga}(1996)}]{seljak96}
{Seljak}, U., \& {Zaldarriaga}, M. 1996, \apj, 469, 437, astro-ph/9603033

\bibitem[{{Seljak} \& {Zaldarriaga}(1999)}]{seljak99}
------. 1999, Physical Review Letters, 82, 2636

\bibitem[{{Shaw} {et~al.}(2010){Shaw}, {Nagai}, {Bhattacharya}, \&
  {Lau}}]{shaw10}
{Shaw}, L.~D., {Nagai}, D., {Bhattacharya}, S., \& {Lau}, E.~T. 2010, \apj,
  725, 1452

\bibitem[{{Shaw} {et~al.}(2009){Shaw}, {Zahn}, {Holder}, \&
  {Dor{\'e}}}]{shaw09}
{Shaw}, L.~D., {Zahn}, O., {Holder}, G.~P., \& {Dor{\'e}}, O. 2009, \apj, 702,
  368

\bibitem[{{Sherwin} \& {Das}(2010)}]{sherwin10}
{Sherwin}, B.~D., \& {Das}, S. 2010, ArXiv e-prints, 1011.4510

\bibitem[{{Shirokoff} {et~al.}(2011){Shirokoff}, {Reichardt}, {Shaw}, {Millea},
  {Ade}, {Aird}, {Benson}, {Bleem}, {Carlstrom}, {Chang}, {Cho}, {Crawford},
  {Crites}, {de Haan}, {Dobbs}, {Dudley}, {George}, {Halverson}, {Holder},
  {Holzapfel}, {Hrubes}, {Joy}, {Keisler}, {Knox}, {Lee}, {Leitch}, {Lueker},
  {Luong-Van}, {McMahon}, {Mehl}, {Meyer}, {Mohr}, {Montroy}, {Padin},
  {Plagge}, {Pryke}, {Ruhl}, {Schaffer}, {Spieler}, {Staniszewski}, {Stark},
  {Story}, {Vanderlinde}, {Vieira}, {Williamson}, \& {Zahn}}]{shirokoff11}
{Shirokoff}, E., {et~al.} 2011, \apj, 736, 61

\bibitem[{{Sievers} {et~al.}(2013){Sievers}, {Hlozek}, {Nolta}, {Acquaviva},
  {Addison}, {Ade}, {Aguirre}, {Amiri}, {Appel}, {Barrientos}, {Battistelli},
  {Battaglia}, {Bond}, {Brown}, {Burger}, {Calabrese}, {Chervenak}, {Crichton},
  {Das}, {Devlin}, {Dicker}, {Bertrand Doriese}, {Dunkley}, {D{\"u}nner},
  {Essinger-Hileman}, {Faber}, {Fisher}, {Fowler}, {Gallardo}, {Gordon},
  {Gralla}, {Hajian}, {Halpern}, {Hasselfield}, {Hern{\'a}ndez-Monteagudo},
  {Hill}, {Hilton}, {Hilton}, {Hincks}, {Holtz}, {Huffenberger}, {Hughes},
  {Hughes}, {Infante}, {Irwin}, {Jacobson}, {Johnstone}, {Baptiste Juin},
  {Kaul}, {Klein}, {Kosowsky}, {Lau}, {Limon}, {Lin}, {Louis}, {Lupton},
  {Marriage}, {Marsden}, {Martocci}, {Mauskopf}, {McLaren}, {Menanteau},
  {Moodley}, {Moseley}, {Netterfield}, {Niemack}, {Page}, {Page}, {Parker},
  {Partridge}, {Plimpton}, {Quintana}, {Reese}, {Reid}, {Rojas}, {Sehgal},
  {Sherwin}, {Schmitt}, {Spergel}, {Staggs}, {Stryzak}, {Swetz}, {Switzer},
  {Thornton}, {Trac}, {Tucker}, {Uehara}, {Visnjic}, {Warne}, {Wilson},
  {Wollack}, {Zhao}, \& {Zuncke}}]{sievers13}
{Sievers}, J.~L., {et~al.} 2013, ArXiv e-prints, 1301.0824

\bibitem[{{Smith} {et~al.}(2009){Smith}, {Senatore}, \&
  {Zaldarriaga}}]{smith09}
{Smith}, K.~M., {Senatore}, L., \& {Zaldarriaga}, M. 2009, Journal of Cosmology
  and Astroparticle Physics, 9, 6

\bibitem[{{Smith} {et~al.}(2007){Smith}, {Zahn}, \& {Dor{\'e}}}]{smith07}
{Smith}, K.~M., {Zahn}, O., \& {Dor{\'e}}, O. 2007, \prd, 76, 043510

\bibitem[{Snyder(1987)}]{snyder87}
Snyder, J.~P. 1987, Map projections--a working manual (Washington: U.S.
  Geological Survey)

\bibitem[{{Song} {et~al.}(2003){Song}, {Cooray}, {Knox}, \&
  {Zaldarriaga}}]{song03}
{Song}, Y.-S., {Cooray}, A., {Knox}, L., \& {Zaldarriaga}, M. 2003, \apj, 590,
  664

\bibitem[{{Story} {et~al.}(2012){Story}, {Reichardt}, {Hou}, {Keisler}, {Aird},
  {Benson}, {Bleem}, {Carlstrom}, {Chang}, {Cho}, {Crawford}, {Crites}, {de
  Haan}, {Dobbs}, {Dudley}, {Follin}, {George}, {Halverson}, {Holder},
  {Holzapfel}, {Hoover}, {Hrubes}, {Joy}, {Knox}, {Lee}, {Leitch}, {Lueker},
  {Luong-Van}, {McMahon}, {Mehl}, {Meyer}, {Millea}, {Mohr}, {Montroy},
  {Padin}, {Plagge}, {Pryke}, {Ruhl}, {Sayre}, {Schaffer}, {Shaw}, {Shirokoff},
  {Spieler}, {Staniszewski}, {Stark}, {van Engelen}, {Vanderlinde}, {Vieira},
  {Williamson}, \& {Zahn}}]{story12}
{Story}, K.~T., {et~al.} 2012, ArXiv e-prints, 1210.7231

\bibitem[{{Sunyaev} \& {Zel'dovich}(1970)}]{sunyaev70}
{Sunyaev}, R.~A., \& {Zel'dovich}, Y.~B. 1970, Comments on Astrophysics and
  Space Physics, 2, 66

\bibitem[{{Sunyaev} \& {Zel'dovich}(1972)}]{sunyaev72}
------. 1972, Comments on Astrophysics and Space Physics, 4, 173

\bibitem[{{Tinker} {et~al.}(2008){Tinker}, {Kravtsov}, {Klypin}, {Abazajian},
  {Warren}, {Yepes}, {Gottl{\"o}ber}, \& {Holz}}]{tinker08}
{Tinker}, J., {Kravtsov}, A.~V., {Klypin}, A., {Abazajian}, K., {Warren}, M.,
  {Yepes}, G., {Gottl{\"o}ber}, S., \& {Holz}, D.~E. 2008, \apj, 688, 709

\bibitem[{{Trac} {et~al.}(2011){Trac}, {Bode}, \& {Ostriker}}]{trac11}
{Trac}, H., {Bode}, P., \& {Ostriker}, J.~P. 2011, \apj, 727, 94

\bibitem[{{van Engelen} {et~al.}(2012){van Engelen}, {Keisler}, {Zahn}, {Aird},
  {Benson}, {Bleem}, {Carlstrom}, {Chang}, {Cho}, {Crawford}, {Crites}, {de
  Haan}, {Dobbs}, {Dudley}, {George}, {Halverson}, {Holder}, {Holzapfel},
  {Hoover}, {Hou}, {Hrubes}, {Joy}, {Knox}, {Lee}, {Leitch}, {Lueker},
  {Luong-Van}, {McMahon}, {Mehl}, {Meyer}, {Millea}, {Mohr}, {Montroy},
  {Natoli}, {Padin}, {Plagge}, {Pryke}, {Reichardt}, {Ruhl}, {Sayre},
  {Schaffer}, {Shaw}, {Shirokoff}, {Spieler}, {Staniszewski}, {Stark}, {Story},
  {Vanderlinde}, {Vieira}, \& {Williamson}}]{vanengelen12}
{van Engelen}, A., {et~al.} 2012, \apj, 756, 142

\bibitem[{{Vanderlinde} {et~al.}(2010){Vanderlinde}, {Crawford}, {de Haan},
  {Dudley}, {Shaw}, {Ade}, {Aird}, {Benson}, {Bleem}, {Brodwin}, {Carlstrom},
  {Chang}, {Crites}, {Desai}, {Dobbs}, {Foley}, {George}, {Gladders}, {Hall},
  {Halverson}, {High}, {Holder}, {Holzapfel}, {Hrubes}, {Joy}, {Keisler},
  {Knox}, {Lee}, {Leitch}, {Loehr}, {Lueker}, {Marrone}, {McMahon}, {Mehl},
  {Meyer}, {Mohr}, {Montroy}, {Ngeow}, {Padin}, {Plagge}, {Pryke}, {Reichardt},
  {Rest}, {Ruel}, {Ruhl}, {Schaffer}, {Shirokoff}, {Song}, {Spieler},
  {Stalder}, {Staniszewski}, {Stark}, {Stubbs}, {van Engelen}, {Vieira},
  {Williamson}, {Yang}, {Zahn}, \& {Zenteno}}]{vanderlinde10}
{Vanderlinde}, K., {et~al.} 2010, \apj, 722, 1180

\bibitem[{{Vieira} {et~al.}(2010){Vieira}, {Crawford}, {Switzer}, {Ade},
  {Aird}, {Ashby}, {Benson}, {Bleem}, {Brodwin}, {Carlstrom}, {Chang}, {Cho},
  {Crites}, {de Haan}, {Dobbs}, {Everett}, {George}, {Gladders}, {Hall},
  {Halverson}, {High}, {Holder}, {Holzapfel}, {Hrubes}, {Joy}, {Keisler},
  {Knox}, {Lee}, {Leitch}, {Lueker}, {Marrone}, {McIntyre}, {McMahon}, {Mehl},
  {Meyer}, {Mohr}, {Montroy}, {Padin}, {Plagge}, {Pryke}, {Reichardt}, {Ruhl},
  {Schaffer}, {Shaw}, {Shirokoff}, {Spieler}, {Stalder}, {Staniszewski},
  {Stark}, {Vanderlinde}, {Walsh}, {Williamson}, {Yang}, {Zahn}, \&
  {Zenteno}}]{vieira10}
{Vieira}, J.~D., {et~al.} 2010, \apj, 719, 763

\bibitem[{{Wilson} {et~al.}(2012){Wilson}, {Sherwin}, {Hill}, {Addison},
  {Battaglia}, {Bond}, {Das}, {Devlin}, {Dunkley}, {D{\"u}nner}, {Fowler},
  {Gralla}, {Hajian}, {Halpern}, {Hilton}, {Hincks}, {Hlozek}, {Huffenberger},
  {Hughes}, {Kosowsky}, {Louis}, {Marriage}, {Marsden}, {Menanteau}, {Moodley},
  {Niemack}, {Nolta}, {Page}, {Partridge}, {Reese}, {Sehgal}, {Sievers},
  {Spergel}, {Staggs}, {Swetz}, {Switzer}, {Trac}, \& {Wollack}}]{wilson12}
{Wilson}, M.~J., {et~al.} 2012, \prd, 86, 122005

\bibitem[{{Zaldarriaga} \& {Seljak}(1999)}]{zaldarriaga99}
{Zaldarriaga}, M., \& {Seljak}, U. 1999, \prd, 59, 123507

\end{thebibliography}

\end{document}